\documentclass[pre,twocolumn]{revtex4-1}

\usepackage{graphicx}
\usepackage{dcolumn}
\usepackage{bm}

\usepackage[utf8]{inputenc}
\usepackage[T1]{fontenc}
\usepackage{mathptmx}

\usepackage{amsmath}
\usepackage{mathtools}
\usepackage{amssymb}

\usepackage{pst-node}
\usepackage{verbatim}   
\usepackage{color}      
\usepackage{subfigure}  
\usepackage{hyperref}   

\usepackage{braket} 

\newcommand{\bnabla}{\mbox{\boldmath $\nabla$}}

\newcommand{\ba}{\begin{eqnarray}}
\newcommand{\ea}{\end{eqnarray}}
\newcommand{\be}{\begin{equation}}
\newcommand{\ee}{\end{equation}}

\pdfoutput=1

\begin{document}

\title{Statistical mechanics of passive Brownian particles in a fluctuating harmonic trap}

\author{Derek Frydel}
\affiliation{Department of Chemistry, Universidad Técnica Federico Santa María, Campus San Joaquin, Santiago, Chile}

\date{\today}

\begin{abstract}
We consider passive Brownian particles trapped in an "imperfect" harmonic trap.  The trap is imperfect 
because it is randomly turned off and on, and as a result particles fail to equilibrate.  Another way to think 
about this is to say that a harmonic trap is time-dependent on account of its strength evolving stochastically in time.  
Particles in such a system are passive and activity arises through external control of a trapping potential, 
thus, no internal energy is used to power particle motion.  A stationary Fokker-Planck equation of this 
system can be represented as a third-order differential equation, and its solution, a stationary distribution, 
can be represented as a superposition of Gaussian distributions for different strengths of a harmonic trap.  
This permits us to interpret a stationary system as a system in equilibrium with quenched disorder.  
\end{abstract}

\pacs{
}

\maketitle

\section{Introduction}

A typical active particle system consists of particles that derive motion from an internal 
energy source.  Standard models that represent such particles are the run-and-tumble (RTP)
\cite{Tailleur08,Tailleur09,Evans18,Dhar18,Dhar19,Basu20,Soto20,Razin20,Frydel21b,Frydel21c,Farago22,Frydel22b,Connor23,Farago24,Ybert24,Loewe24} 
and active-Brownian particle (ABP) models \cite{Dhar20,Cargalio22}, one more suitable to 
biological and the other to chemical systems.  Another active scenario arises when a 
passive Brownian particle is immersed in a bath of active particles.  The model that 
addresses this situation is the active Ornstein-Uhlenbeck particle model (AOUP)
\cite{Szamel14,Shankar18,Sevilla19,Cates21,Marconi18,Lowen22}.  

A different and somewhat less explored design of active dynamics is realized by placing passive Brownian 
particles in a fluctuating external potential, thereby preventing particles from equilibrating 
\cite{Pal2013,Wang17,Santra21,Kundu21,Kundu21b,Cocconi22}.  
In this 
setup, there is no need for a special type of particles and the only requirement is that an external 
potential varies stochastically in time.  A possible experimental realization of such a system 
could be attained using tweezer instruments and techniques 
\cite{Yael11,Brady16,Brady21,Lowen22b,Genet22,Yael23}.  


This work considers passive Brownian particles trapped in a harmonic potential $u = Kr^2/2$
with time-dependent strength $K\equiv K(t)$.  We consider a specific evolution in which 
$K(t)$ changes discontinuously between two discrete values, so the trap is either in an off or on 
state.  
The amount of time a particle spends in each state is drawn from an exponential distribution.  
%

A stationary state of this system is governed by a third-order differential equation.   
A third-order differential equation were previously found to describe a stationary state of run-and-tumble particles 
in a static harmonic trap in one- and two-dimensions \cite{Frydel22c,Frydel23b,Frydel24a}, for which 
a solution was found to be a convolution between two distributions.  
A stationary solution of the present system can be represented as a superposition of Gaussian 
distributions for different strengths of a harmonic trap.  This permits us to interpret a stationary state 
as a system in equilibrium with quenched disorder.  
The superposition of Gaussian distributions has been encountered in quantum theory and financial markets 
\cite{Kleinert04,Kleinert08}. In non-equilibrium statistical mechanics, those distributions have
 been considered in \cite{Beck01} in the context of non-extensive statistical mechanics. 


The current work could be viewed in relation to systems that explore control of external forces 
and information feedback to perform work \cite{Sagawal15,Sagawa08,Cao09,Horowitz10,Sagawa10,Pon10,Suzuki10,Kundu12}.  
Although the model that is analyzed does not involve information feedback and 
the only control comes from regulating the time duration when particles are either trapped or released, it could be 
considered a small step in theoretical understanding of this class of problems. 

Previous work done on systems with fluctuating external potentials include the study of particles 
in a harmonic trap with a fluctuating trap center \cite{Pal2013} that was motivated by an 
experiment \cite{Gomez2010}.  A system of particles trapped in a harmonic
trap with a fluctuating strength was first investigated in \cite{Santra21}.   Particles in a fluctuating 
linear potential were investigated in \cite{Kundu21,Kundu21b} as a species of a resetting problem.    
The entropy production rate of single particle inside a fluctuating external 
potential of an arbitrary shape was calculated in \cite{Cocconi22}.  





This work is organized as follows. In Sec. (\ref{sec:model})  we introduce the model and 
obtain a third-order differential equation for a stationary state. 
In Sec. (\ref{sec:superposition}) we represent the solution as a superposition 
of Gaussian distributions for different trap strengths. 
 In Sec. (\ref{sec:time})  we consider 
particles trapped in a harmonic potential with a general time-dependent trap strength.  It is 
shown that the time-dependent distribution at any given moment is a Gaussian distribution. 
This result justifies the use of the superposition formula. In Sec. (\ref{sec:extension}) 
we extend all the exact results to a system for an arbitrary dimension. 
In Sec. (\ref{sec:quantities}) 
we consider quantities of physical interest.  The work is 
concluded in Sec. (\ref{sec:conclusion}).

\section{The model }
\label{sec:model}

The focus of this work is a conceptually simple model:   
an ideal gas trapped in an "imperfect" harmonic potential.  The potential 
is imperfect because it is turned off and on at random time intervals.  The alternate cycle of trapping and 
releasing prevents particles from attaining equilibrium and as a result gives rise to a non-equilibrium situation.   
The times $t_p$ during which a particle is either trapped 
or released are sampled from the exponential distribution $\sim e^{- t_p/\tau}$, where $\tau$ is the average 
time during which a particle persists in a given state.  At the end of each time $t_p$, a particle switches 
to another state with probability one.  

Since the times when a trap is either in the "on" or "off" state are sampled from the same distribution, the 
system spends on average the same amount of time in both states.   What changes is the rate, given 
by $\tau^{-1}$, with which the trap fluctuates between the two states.  The model represents a specific 
case of a harmonic potential $u=Kr^2/2$ with the time-dependent strength $K\equiv K(t)$. 

For a system in one-dimension, the Fokker-Planck formulation that describes such a model might be written 
as
\ba
&&    \dot \rho_+      =       \mu \left( \frac{K}{2} + \Delta\right)  [x\rho_+]'      +     D\rho_+''      +      \frac{1}{\tau}      \left( \rho_-   -  \rho_+ \right)    \nonumber\\
&&    \dot \rho_-       =       \mu \left( \frac{K}{2}  -  \Delta\right)  [x\rho_-]'     +     D\rho_-''       +      \frac{1}{\tau}      \left( \rho_+  -  \rho_- \right), 
\label{eq:FP0}
\ea
where $\mu$ is the mobility, $D=\mu k_BT$ is the diffusion constant, $T$ is the temperature, and 
$k_B$ is the Boltzmann constant.  To simplify the expressions, we use the dot notation to represent the 
time derivative, $\dot \rho_+\equiv \frac{\partial \rho_+}{\partial t}$, and the prime notation to represent 
derivatives with respect to position, 
$[x\rho_{+}]'  \equiv \frac{\partial }{\partial x}[x\rho_+]$ and $\rho_+'' \equiv \frac{\partial^2 }{\partial x^2}$.  
$\rho_{+}$ and $\rho_{-}$ are the distributions of particles in a harmonic potential with the 
respective strength $K/2 + \Delta$ and $K/2 - \Delta$.  The last term on the right-hand side of each 
equation represents the conversion of one type of particle into another and provides coupling between 
the two equations and also prevents particles to equilibrate.  
The remaining terms represent the usual flux given by 
$$
j_{\pm}     =      -\mu ( K/2 \pm \Delta) x \rho_{\pm}       -       D\rho_{\pm}'.
$$


In this work we are interested in the situation $\Delta = K/2$, in which case the Fokker-Planck equation becomes \cite{Santra21}
\ba
&&    \dot \rho_+      =       \mu K [x\rho_+]'      +     D\rho_+''      +      \frac{1}{\tau}      \left( \rho_-   -  \rho_+ \right)    \nonumber\\
&&    \dot \rho_-       =                                             D\rho_-''       +      \frac{1}{\tau}      \left( \rho_+  -  \rho_- \right).  
\label{eq:FP}
\ea
The distribution $\rho_-$ represents unconstrained particles corresponding to the trap being turned off, 
and $\rho_+$ represents particles subject to a confining potential.  

To simplify the notation, we introduce dimensionless parameters.  
The position of particles on the $x$-axis 
and the rate at which a harmonic trap fluctuates in dimensionless units are then given by 
$$
z = x \sqrt{\frac{\mu K}{D}}, ~~~~~~~~ \alpha = \frac{1}{\tau \mu K}.  
$$
At a stationary state and dimensionless parameters Eq. (\ref{eq:FP}) becomes
\ba
&&   0     =        [z\rho_+]'  +    \rho_+''      +      \alpha    \left( \rho_-   -  \rho_+ \right)    \nonumber\\
&&   0      =      \rho_-''      +      \alpha   \left( \rho_+  -  \rho_- \right).  
\label{eq:FPS}
\ea

The two equations can be merged into a single differential equation for a distribution $\rho = (\rho_+  +  \rho_-)/2$. 
The result is a third-order differential equation 
\be
0      =         -  \alpha  z^3 \rho    +      2( 1    -   \alpha z^2 ) \rho'    -    ( 2 - z^2 )  z \rho''    +      z^2 \rho'''.   
\label{eq:diff-3}
\ee
See Appendix (\ref{app:sec1}) for details.  

In the limit of a fast appearing-disappearing trap, $\alpha\to\infty$, Eq. (\ref{eq:diff-3})
reduces to 
\be
0      =     - \frac{z\rho}{2}     -    \rho',
\label{eq:diff-3-lim}
\ee
where $\rho$ is a Gaussian function $\sim e^{-z^2/4}$, which can be identified with passive particles in equilibrium 
but in a harmonic trap that is half the strength of the original trap.   


Finding a solution to Eq. (\ref{eq:diff-3}) for an arbitrary $\alpha$ is more challenging.  While first- 
and second-order differential equations are relatively common in physics, third-order differential 
equations are encountered less frequently \cite{Pati13}.  They are also more challenging to solve 
and analyze.  
It is relatively easy, however, to analyze Eq. (\ref{eq:diff-3}) by analyzing its moments.  

The moments of a stationary distribution $\rho$ can be obtained 
by transforming Eq. (\ref{eq:diff-3}) into a recurrence relation by operating on it with 
$\int_{-\infty}^{\infty}dz\, z^{2n-3}$ \cite{Frydel22c,Frydel23b,Frydel24a}.  Followed by the integration by parts this yields 
\be
m_{n}   =    \frac{4n - 2}{\alpha}  \bigg[  ( n  +  \alpha )  m_{n-1}    -    n ( 2n - 3 )  m_{n-2} \bigg], 
\ee
where $m_{n} = \langle z^{2n} \rangle$ are even moments of $\rho$, all odd moments being zero.  

Given the initial condition $m_0=1$, all subsequent moments can be calculated, and the first two terms of 
the sequence are
\ba
&& \langle z^{2} \rangle     =       \left(  2   +   \frac{1}{ \alpha}  \right) \nonumber\\
&& \langle z^{4} \rangle  =   6 \left( 2 +  \frac{3}{\alpha}   +    \frac{2}{\alpha^2} \right).  
\label{eq:m}
\ea
In the limit $\alpha\to \infty$, all the moments reduce to the moments of a Gaussian distribution $\sim e^{-z^2/4}$.  
The moments increase as $\alpha$ decreases, indicating the spreading out of $\rho$.

Moments of distributions corresponding do a different state can be calculated by operating on 
the two equations in Eq. (\ref{eq:FPS}) with $\int dz\, z^{2n+2}$.  After some manipulation, 
the details of which can be found in Appendix \ref{app:sec2a}, we get two coupled recurrence relations, 
from which we get
\ba
&& \langle z^{2} \rangle_+     =       2  \nonumber\\
&& \langle z^{4} \rangle_+  =   6 \left(2   +  \frac{1} {\alpha} \right),
\label{eq:mp}
\ea
for particles in a trapped state, and 
\ba
&&  \langle z^{2} \rangle_-     =       2 \left(1  +    \frac{1}{\alpha}\right)  \nonumber\\
&&  \langle z^{4} \rangle_-  =   6\left( 2 +  \frac{5  } {\alpha} +  \frac{4 } {\alpha^2} \right), 
\label{eq:mn}
\ea
 for particles in a released state.  
An interesting observation is that the second moment of trapped particles does not 
depend on $\alpha$, $\langle z^{2} \rangle_+ = 2$.  
This feature of a system will have interesting repercussions on various physical quantities 
discussed later in this work.

\section{Solution as a superposition of Gaussian distributions}
\label{sec:superposition}

In this section we develop a method that would permit us to solve a third-order
differential equation in Eq. (\ref{eq:diff-3}). 
 As mentioned earlier, third-order differential equations are less common in physics.  
In the context of active particles, a third-order differential equation arises for RTP particles 
in a static harmonic trap in one- and two-dimensions \cite{Frydel22c}.  For higher dimensions, 
a differential equation of the same system has a more complex structure and to date it 
was not possible to obtain it.    


To solve Eq. (\ref{eq:diff-3}) we represent $\rho$ as a superposition of Gaussian distributions for 
different effective strengths, 
\be
\rho(z)   =   \int_{0}^{1} d\lambda \, p(\lambda)    \rho_G(z;\lambda), 
\label{eq:rho-p}
\ee
where 
\be
\rho_G(z;\lambda) = \sqrt{\frac{\lambda}{2\pi}} e^{-\lambda z^2/2}
\label{eq:rhoG}
\ee
is a Gaussian distribution for a dimensionless strength $\lambda$.  By representing the stationary 
distribution $\rho$ as a superposition of equilibrium distributions (since $\rho_G$ is an equilibrium 
distribution for an effective strength $\lambda$), we reinterpret our system as not out-of-equilibrium 
but as effectively at equilibrium with quenched disorder.  Thus, whatever features of $\rho$ that make it
deviate from a Gaussian form can be attributed to quenched disorder.  


Other than providing an alternative interpretation of the original system, there is a certain mathematical 
advantage of representing a stationary distribution as a superposition formula, as this shifts the task 
from finding $\rho$ to that of finding $p(\lambda)$, where it is reasonable to assume that $p$ is less 
complex than $\rho$. 


We note that since $p(\lambda)$ is normalized, it follows that $\int_{-\infty}^{\infty} dz\, \rho = 1$.  
Also, $p(\lambda)$ is defined on $\lambda\in [0,1]$, since 
contributions $\lambda>1$ are not physically meaningful given that the actual 
system switches between the states $\lambda=0$ and $\lambda=1$.  This means that the superposition 
should span a continuous region between those two states, since the distribution itself does 
not jump from one state to another but evolves between the two states in a continuous manner.

To proceed, we distinguish between distributions 
$p_+(\lambda)$ and $p_-(\lambda)$, such that $\rho_{\pm} = \int_0^{1} d\lambda\, p_{\pm}(\lambda) \rho_G(z;\lambda)$. 
We then infer a differential equation for each distribtion from Eq. (\ref{eq:FPS}).  We use the word "infer" intentionally 
as there is no standard technique for obtaining a differential equation for $p_{\pm}$ from Eq. (\ref{eq:FPS}). 
Instead, one proceeds using a trial and error approach.  The two differential equations that were determined are
\ba
&& 0 =        \frac{\partial}{\partial\lambda} [\lambda (\lambda  -  1)  p_+]     +     \frac{\alpha}{2}   ( p_-  -  p_+ )   \nonumber\\
&& 0 =        \frac{\partial}{\partial\lambda} [\lambda^2 p_-]        +         \frac{\alpha} {2} ( p_+  -  p_- ).
\label{eq:FP-p}
\ea
Operating on each of the above equations with $\int_{0}^{1} d\lambda \,  \rho_G(z;\lambda)$ recovers
Eq. (\ref{eq:FPS}), which confirms the accuracy of Eq. (\ref{eq:FP-p}).   
What is important about the result in Eq. (\ref{eq:FP-p}) is that both equations are of a lower order 
than Eq. (\ref{eq:FPS}).  
Merging the two equations in Eq. (\ref{eq:FP-p}) leads to 
\be
0    =     \left( 3 \lambda^2 -  2 \lambda  -    \alpha \lambda  +  \frac{\alpha}{2} \right) p      +    \left( \lambda  -  1 \right) \lambda^2 p', 
\label{eq:diff-p-1}
\ee
where $p = (p_+ + p_-)/2$.  See Appendix (\ref{app:sec2}) for details.  
Eq. (\ref{eq:diff-p-1}) is a first-order differential equation that can be easily solved.  The solution is 
\be
p(\lambda)   =  \frac{1}{2} \frac{ (\alpha/2)^{\alpha/2} e^{\alpha/2} }{ \Gamma(\alpha/2) }   \left( \frac{1}{\lambda} - 1\right)^{\frac{\alpha}{2} - 1}  
\frac{e^{-\frac{\alpha}{2} \frac{1}{\lambda} }}{\lambda^3}, 
\label{eq:p-lambda}
\ee
and the solution for separate distributions $p_{\pm}$ are  
\ba
&& p_+ = 2\lambda p \nonumber\\
&& p_- = 2(1-\lambda) p, 
\label{eq:ppm}
\ea
with $p$ given in Eq. (\ref{eq:p-lambda}).  Both solutions can be verified by inserting into Eq. (\ref{eq:FP-p}).

\begin{figure}[hhhh] 
 \begin{center}
 \begin{tabular}{rrrr}
 \includegraphics[height=0.19\textwidth,width=0.21\textwidth]{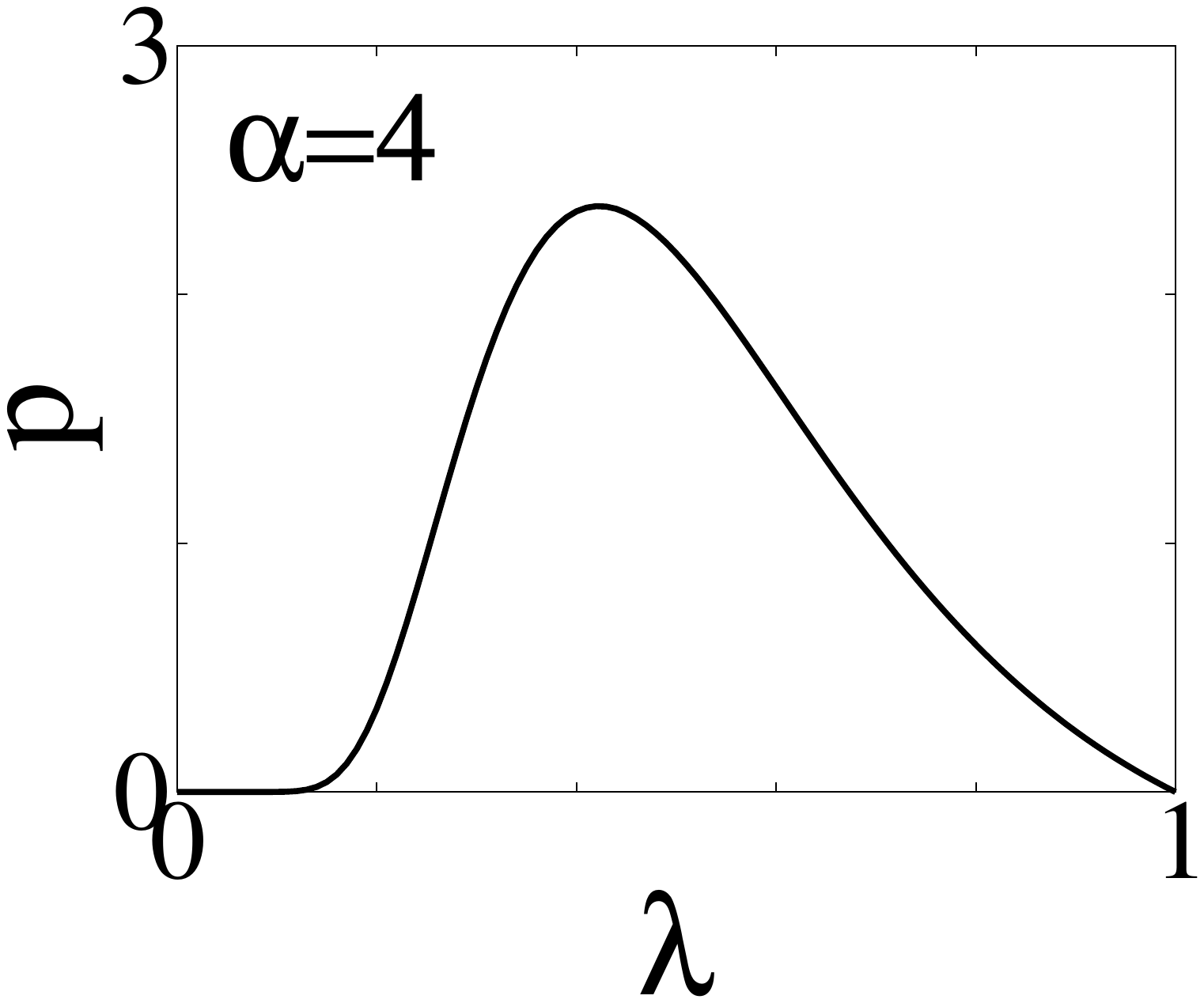} &
 \includegraphics[height=0.19\textwidth,width=0.21\textwidth]{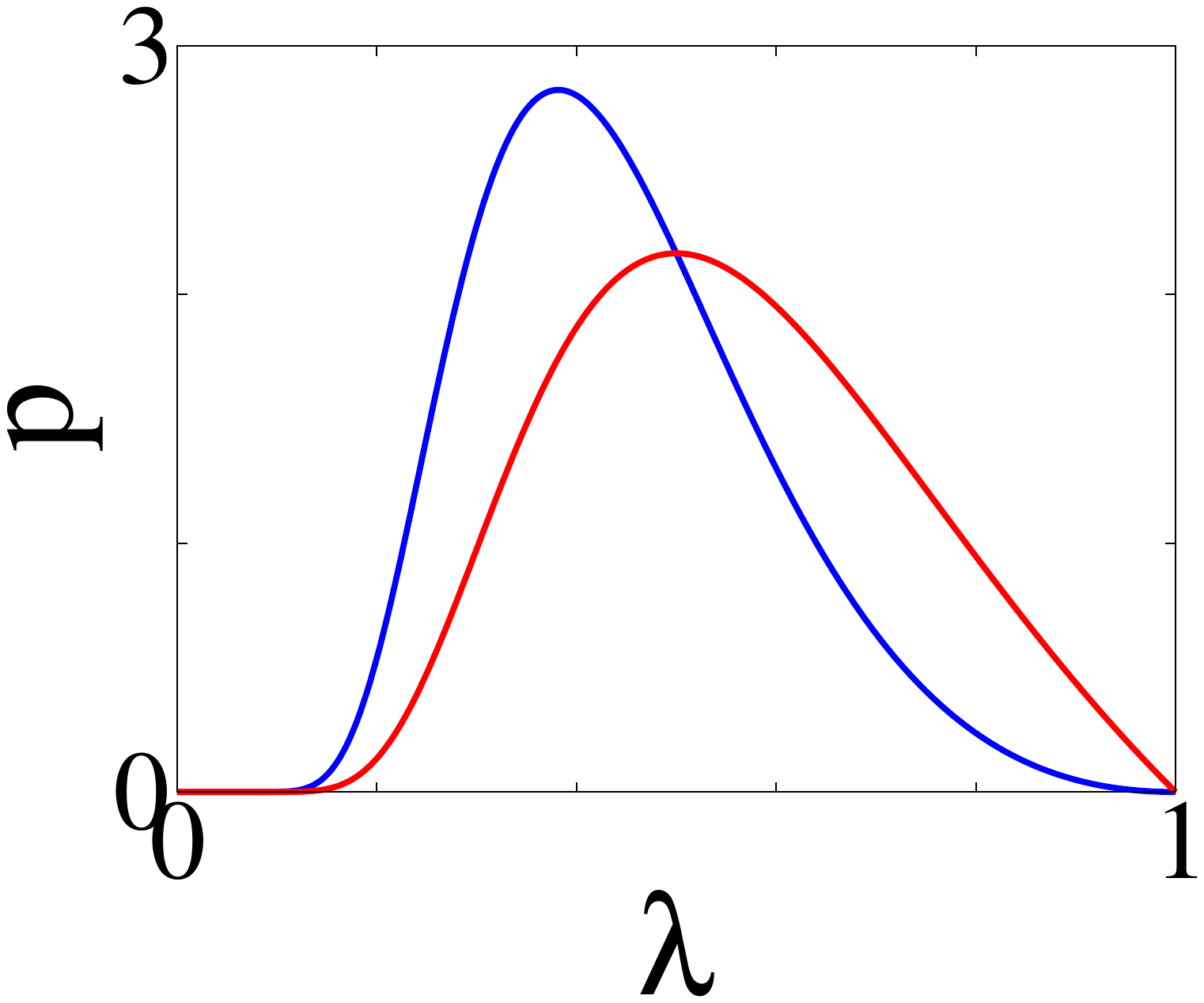} \\
 \includegraphics[height=0.19\textwidth,width=0.21\textwidth]{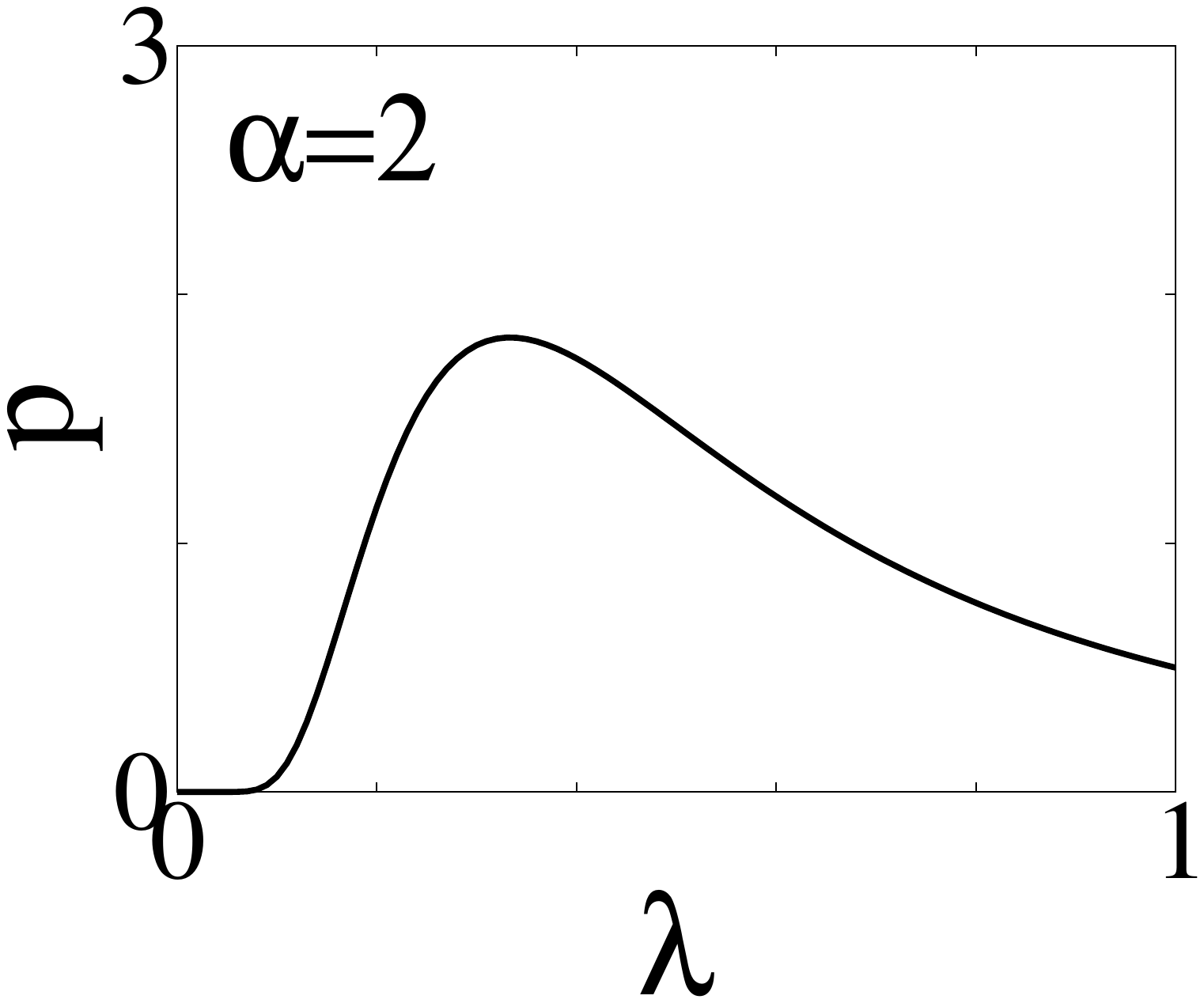} &
  \includegraphics[height=0.19\textwidth,width=0.21\textwidth]{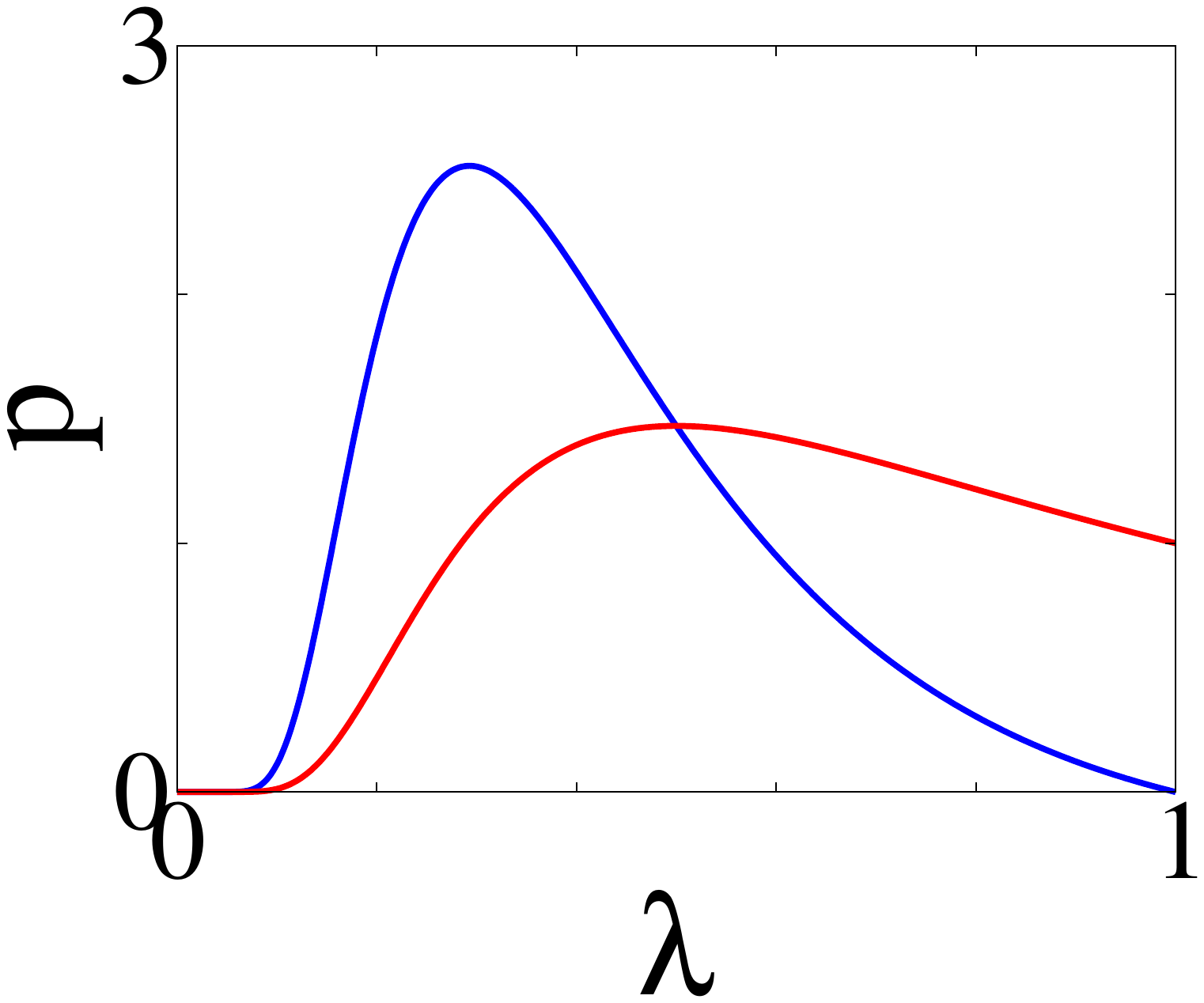} \\
\includegraphics[height=0.19\textwidth,width=0.21\textwidth]{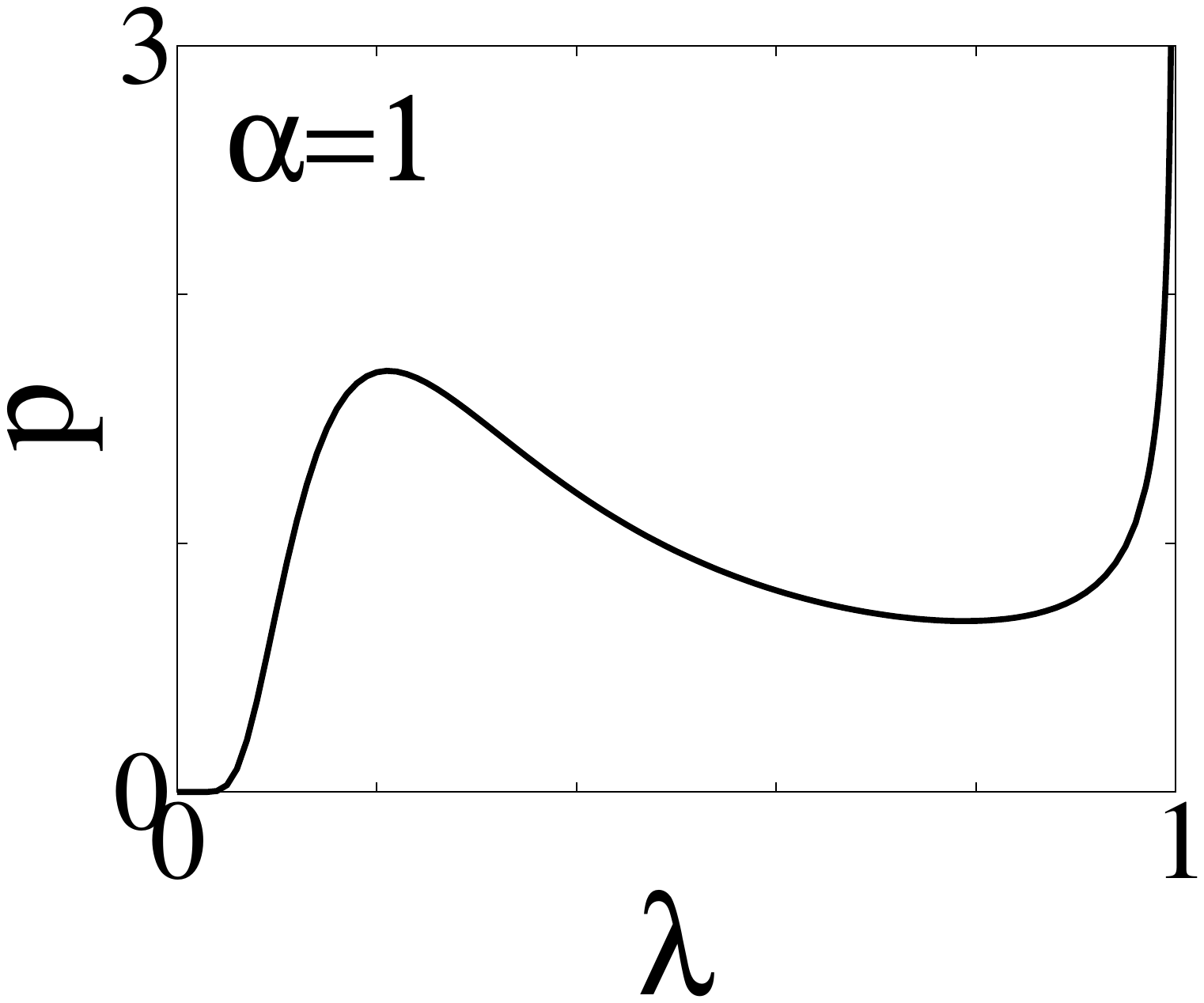} &
\includegraphics[height=0.19\textwidth,width=0.21\textwidth]{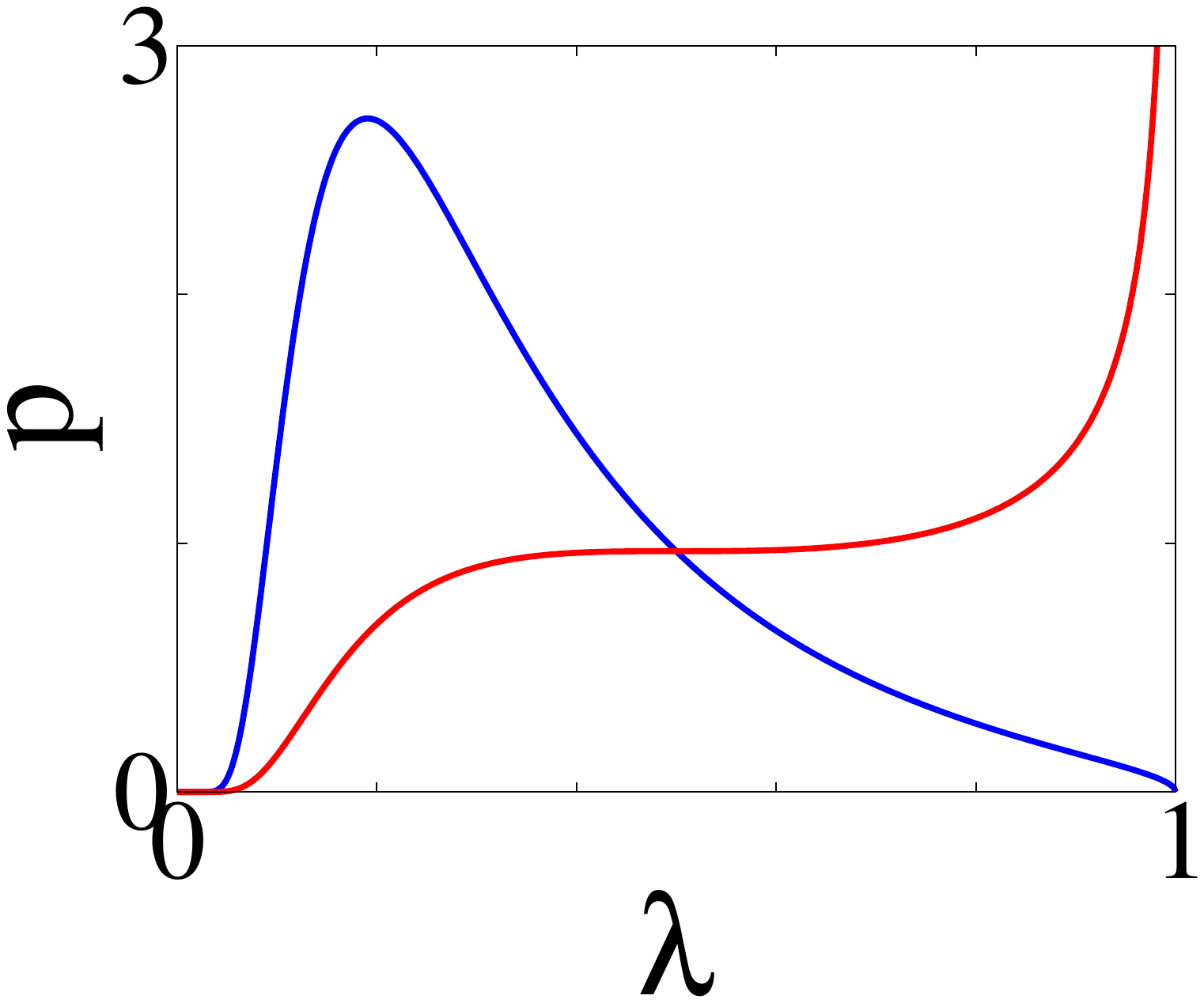} &
\end{tabular}
 \end{center} 
\caption{ Probability distribution $p(\lambda)$ given in Eq. (\ref{eq:p-lambda}) for different values of 
$\alpha$.   Distributions on the right-hand side column 
are for $p_+$ and $p_-$.  Blue colored distributions represent released 
particles $p_-$ and red colored distributions represent trapped particles $p_+$.  
} 
\label{fig:p} 
\end{figure}

Examining the expression in Eq. (\ref{eq:p-lambda}) we note that for $\alpha=2$ the factor  
$\left( {1} / {\lambda} - 1\right)^{\frac{\alpha}{2} - 1}$ reduces to unity, indicating some sort of 
crossover for this value of $\alpha$.
To better see how this crossover manifests itself, in Fig. (\ref{fig:p}) we plot $p(\lambda)$, 
given in Eq. (\ref{eq:p-lambda}), for different values of $\alpha$.  
\graphicspath{{figures/}}
We recall that $p(\lambda)$ is defined on $\lambda\in [0,1]$, as the system fluctuates 
between the states $\lambda=0$ and $\lambda=1$.  
The fact that $p(0)=0$ implies that the sate $\lambda=0$ is not really accessible since for 
any finite $\alpha$ particles remain confined to a finite region.  To reach the state $\lambda=0$
the system would require an infinite amount of time.

The behavior at $\lambda=1$ is more interesting. 
The probability at $\lambda=1$ exhibits a crossover at $\alpha=2$ such that 
\be
p(1) 
=
\begin{cases}
0 & \text{$\alpha>2$} \\
\frac{1}{2}  & \text{$\alpha=2$} \\
\infty  & \text{$\alpha<2$}.  
\label{eq:cross}
\end{cases}
\ee
If fluctuations of a trap strength are rapid enough, such that $\alpha>2$, then
the state $\lambda=1$ is never attained and $p(1)=0$.  
For a slower rate of fluctuations, such as $\alpha<2$, particles remain trapped 
for a sufficiently long time.  This permits a system to come very close to $\lambda=1$, 
giving rise to a divergence at $\lambda=1$.    The divergence, however, is integrable, 
indicating that the state $\lambda=1$ is being approached but never truly attained.  

Note that although we analyze a stationary distribution, we talk about a system as evolving between two states.  
The system is stationary when it is averaged over time, or over many independent ensembles. 
We will discuss the aspect of time dependence in more detail in Sec. (\ref{sec:time}).

Separate distributions $p_{\pm}(\lambda)$ are displayed on the right-hand-side in Fig. (\ref{fig:p}).  
The distributions $p_+$ and $p_-$ become increasingly similar with increasing $\alpha$ and 
converge in the limit $\alpha\to\infty$.  On the other hand, the two distributions become increasingly 
distinct as $\alpha$ becomes smaller.  The two distributions shift toward a different side of the domain and 
evolve into very different shapes.  The released particles move toward the state $\lambda=0$ and 
the trapped particles concentrate around $\lambda=1$.  

\begin{figure}[hhhh] 
 \begin{center}
 \begin{tabular}{rrrr}
 \includegraphics[height=0.19\textwidth,width=0.21\textwidth]{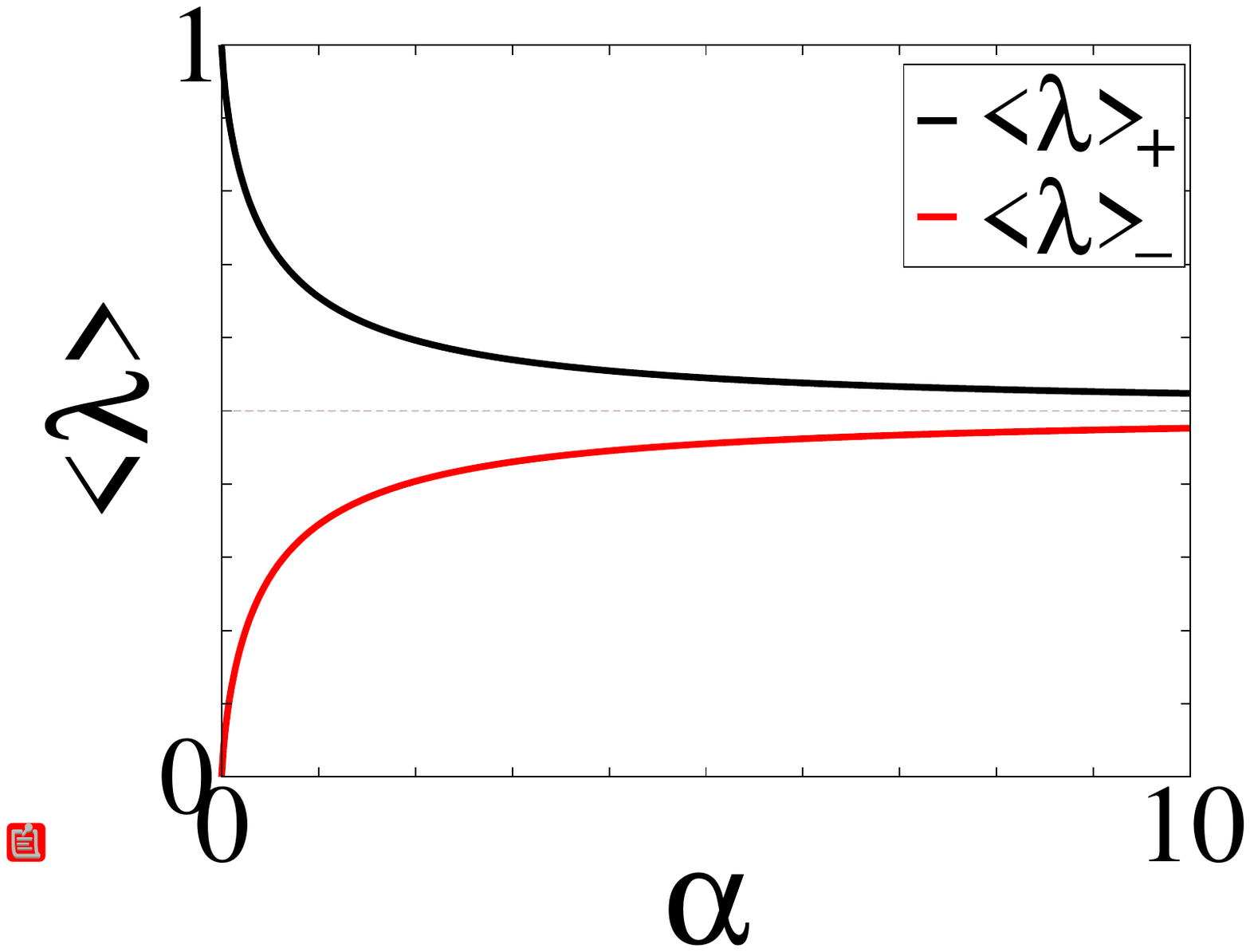} &
\end{tabular}
 \end{center} 
\caption{ The first moment of $p_+(\lambda)$ and $p_{-}(\lambda)$ as a function of $\alpha$.  } 
\label{fig:ave-l} 
\end{figure}

Despite very different shapes, the first moment of each distribution, shown 
in Fig. (\ref{fig:ave-l}), are symmetric around $\lambda=1/2$, and as a result 
\graphicspath{{figures/}}
the average value of $\lambda$ of the total distribution $p = (p_+ + p_-)/2$ is fixed at 
$\langle \lambda \rangle  =  \frac{1}{2}$ and is independent of $\alpha$.  There is no 
symmetry between higher moments.  


Using the superposition formula in Eq. (\ref{eq:rho-p}) together with the formula for $p(\lambda)$ 
in Eq. (\ref{eq:p-lambda}), the solution of the third-order differential equation in Eq. (\ref{eq:diff-3}) 
can be expressed as 
\be
\rho(z)   =   \frac{ ( \alpha e /2)^{\alpha/2} }{ 2\Gamma(\alpha/2) }  \left(\frac{1}{2\pi } \right)^{\frac{1}{2}}   \int_0^1 d\lambda\,   
\left( \frac{1}{\lambda} - 1\right)^{\frac{\alpha}{2} - 1}  \frac{e^{-\frac{\alpha}{2\lambda}} e^{-\frac{\lambda  z^2}{2}} } { \lambda^{3-\frac{1}{2}} }.
\label{eq:rho-full}
\ee
The integral in Eq. (\ref{eq:rho-full}) can be evaluated exactly for specific values of $\alpha$ 
corresponding to $\alpha=2n$ where $n$ is the positive integer \cite{Santra21}.  
For $\alpha=2$, which according to 
Eq. (\ref{eq:cross}) corresponds to a crossover, the term $\left( \frac{1}{\lambda} - 1\right)^{\frac{\alpha}{2} - 1}$ 
in Eq. (\ref{eq:rho-full}) becomes unity and $p(\lambda)$ evaluates to the following analytical form
\ba
\rho &=& \frac{1}{2} \frac{e^{-z^2/2}}{\sqrt{2\pi}}  \nonumber\\ 
&+&  \frac{1}{8\sqrt{2}} \left[ \left( 1-\sqrt{2z^2} \right)  \text{erfc}\left[ 1 + \frac{\sqrt{2z^2}}{2} \right]  e^{1+\sqrt{2z^2}} \right] \nonumber\\
&+&  \frac{1}{8\sqrt{2}} \left[ \left( 1+\sqrt{2z^2} \right)  \text{erfc}\left[ 1 - \frac{\sqrt{2z^2}}{2} \right]  e^{1-\sqrt{2z^2}} \right]. 
\label{eq:rho-full-b4}
\ea

\begin{figure}[hhhh] 
 \begin{center}
 \begin{tabular}{rrrr}
 \includegraphics[height=0.19\textwidth,width=0.21\textwidth]{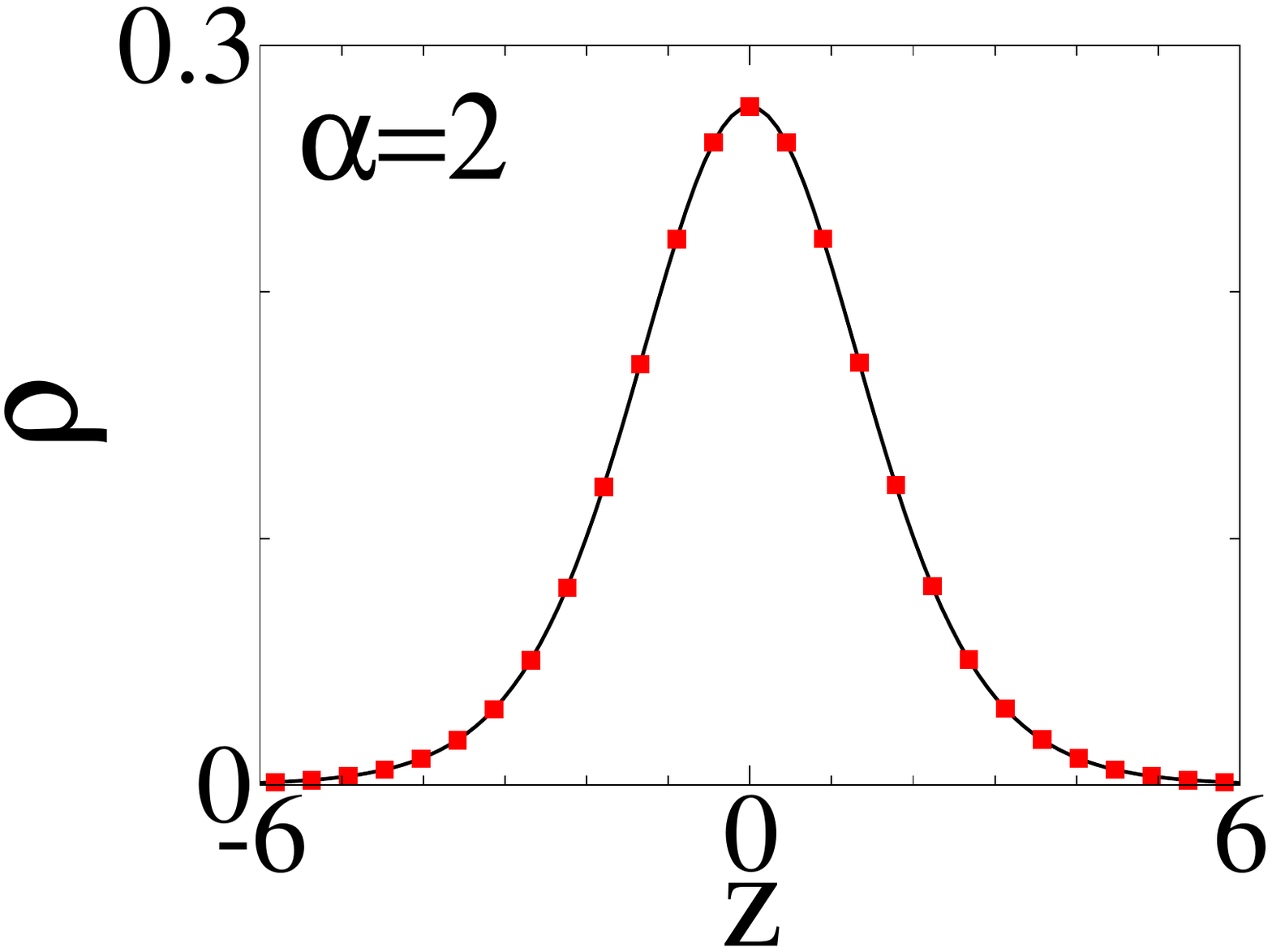} \\
 \includegraphics[height=0.19\textwidth,width=0.21\textwidth]{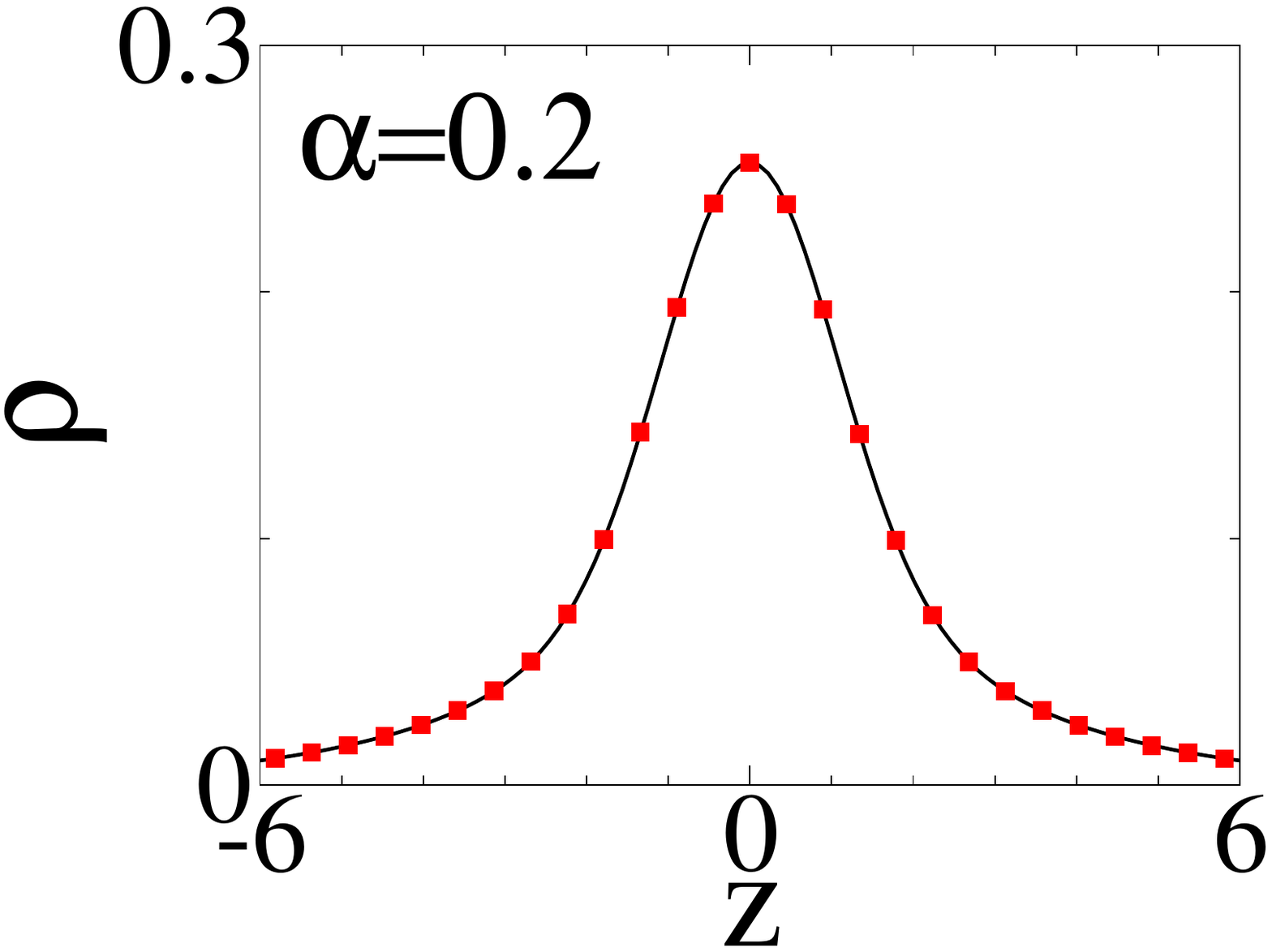} \\
 \includegraphics[height=0.19\textwidth,width=0.21\textwidth]{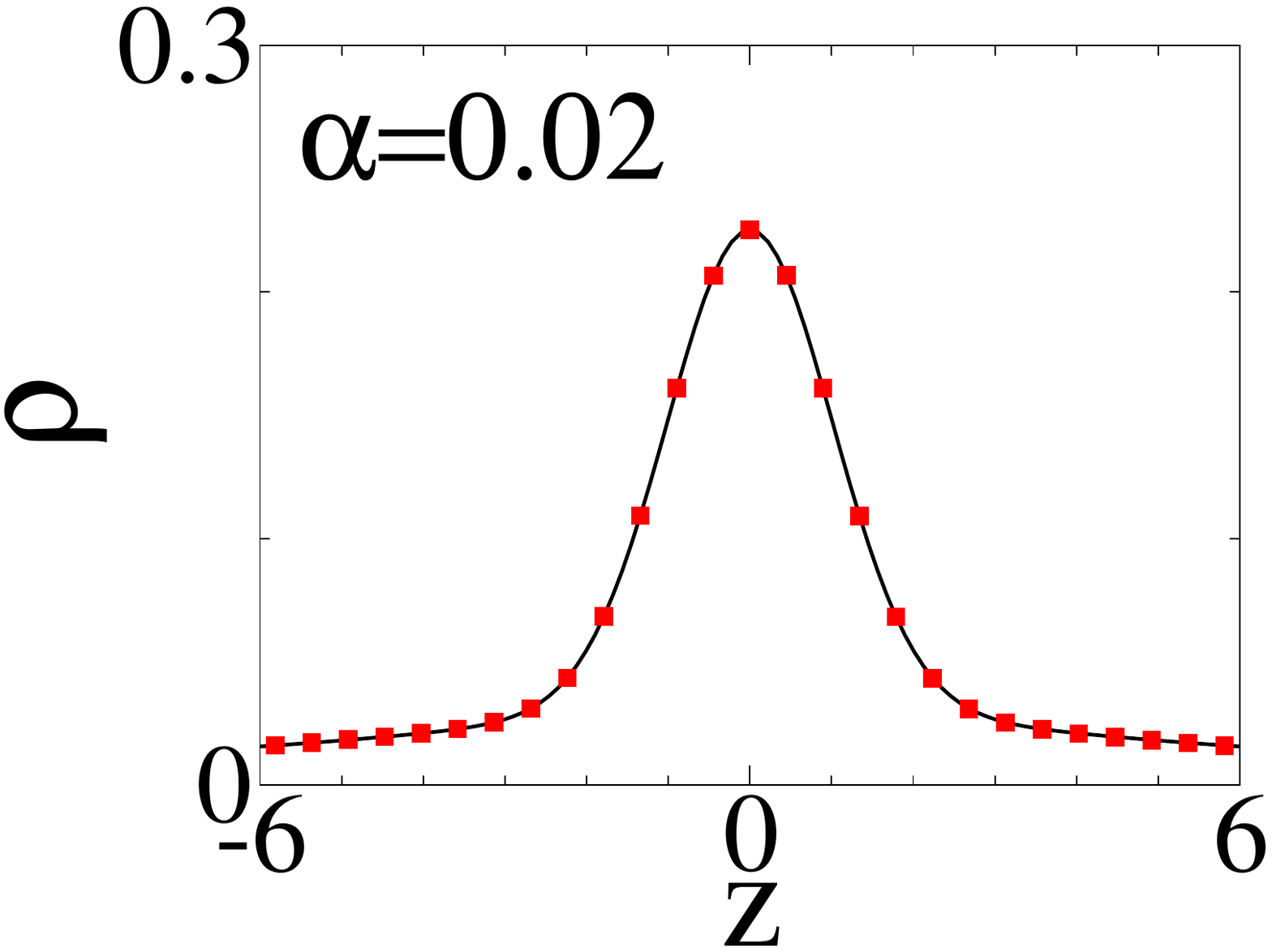} 
\end{tabular}
 \end{center} 
\caption{ Distributions $\rho(z)$ calculated using Eq. (\ref{eq:rho-full}) for different values of 
$\alpha$.   Red data points are obtained from a simulation 
based on numerical integration of the Langevin equation.   } 
\label{fig:rho} 
\end{figure}

In Fig. (\ref{fig:rho}) we plot $\rho(z)$ for different values of $\alpha$.  For those values of $\alpha$ where 
no analytical expression is available, the integral in Eq. (\ref{eq:rho-full}) is evaluated numerically.   All probability 
distributions are in addition compared with distributions obtained from simulation based on numerical integration 
of the Langevin equation to confirm the correctness of analytical results.  
\graphicspath{{figures/}}
The Langevin equation is integrated using the Euler method, 
\ba
&& x_+(t+\Delta t) = x_+(t) - \mu K x_+(t) + \xi_+(t) \sqrt{2D\Delta t} \nonumber\\
&& x_-(t+\Delta t) = x_-(t) + \xi_-(t) \sqrt{2D\Delta t} 
\ea
where $\xi_{\pm}(t)$ is the white noise with zero mean and unity variance, and $x_+$
and $x_-$ is the position of a particle in a trapped and released state, respectively.  
An individual particle changes from $x_+$ to $x_-$ and vice versa at the end of the 
time $t_p$ drawn from the exponential distribution $\sim e^{-t_p/\tau}$.

The primary non-Gaussian feature of the distributions is that particles spread out beyond the trap
boundaries.  This feature becomes more pronounced 
for small values of $\alpha$ where particles are permitted to remain in a given state for a longer 
time.  By allowing particles to be released for a longer time, particles become more spread out.  What 
is interesting is that we did not observe any distinct features in $\rho$ that would signal the crossover 
at $\alpha=2$.  The crossover appears to be primarily the feature of $p(\lambda)$.     

\begin{figure}[hhhh] 
 \begin{center}
 \begin{tabular}{rrrr}
 \includegraphics[height=0.19\textwidth,width=0.21\textwidth]{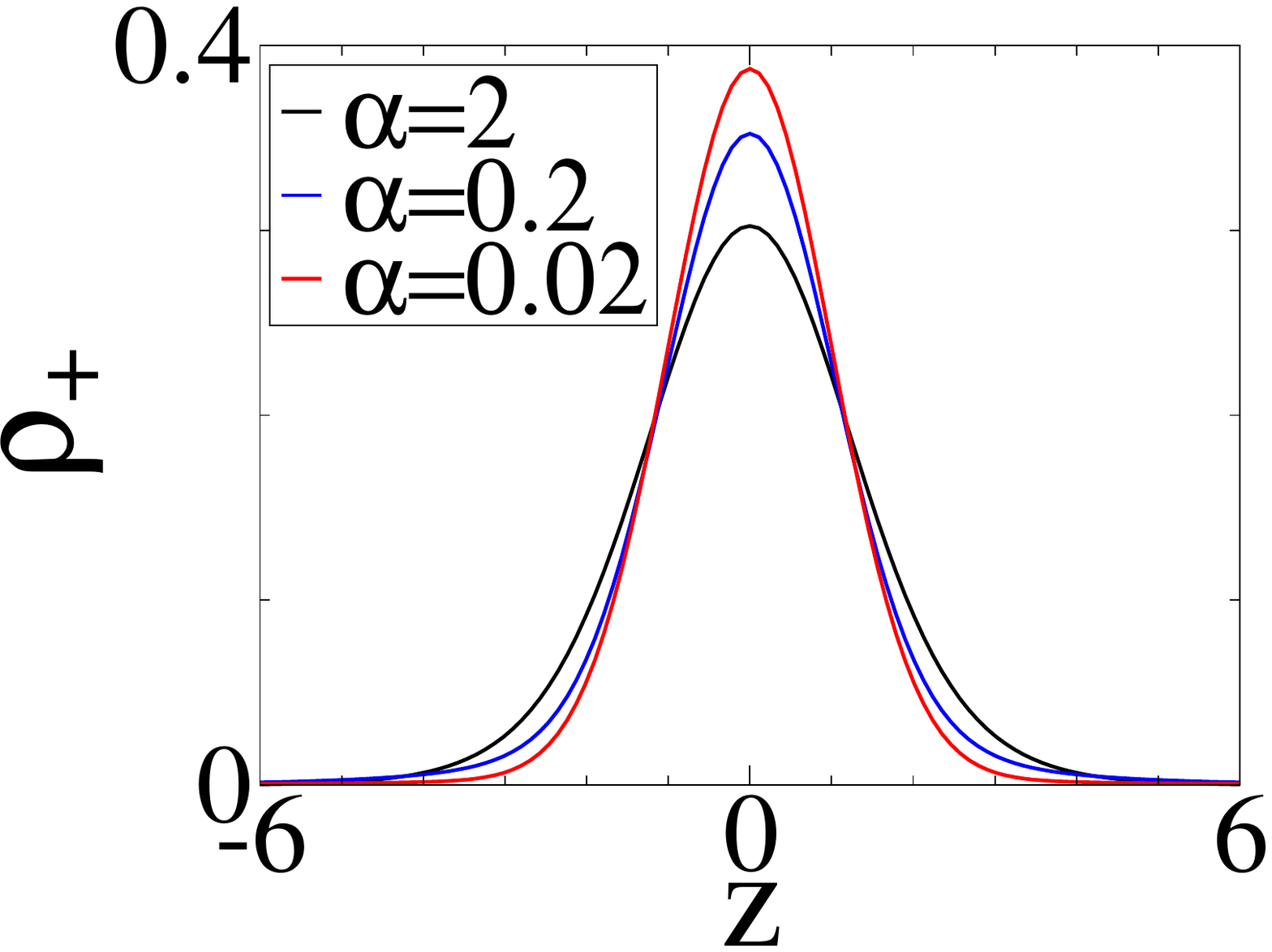} &
 \includegraphics[height=0.19\textwidth,width=0.21\textwidth]{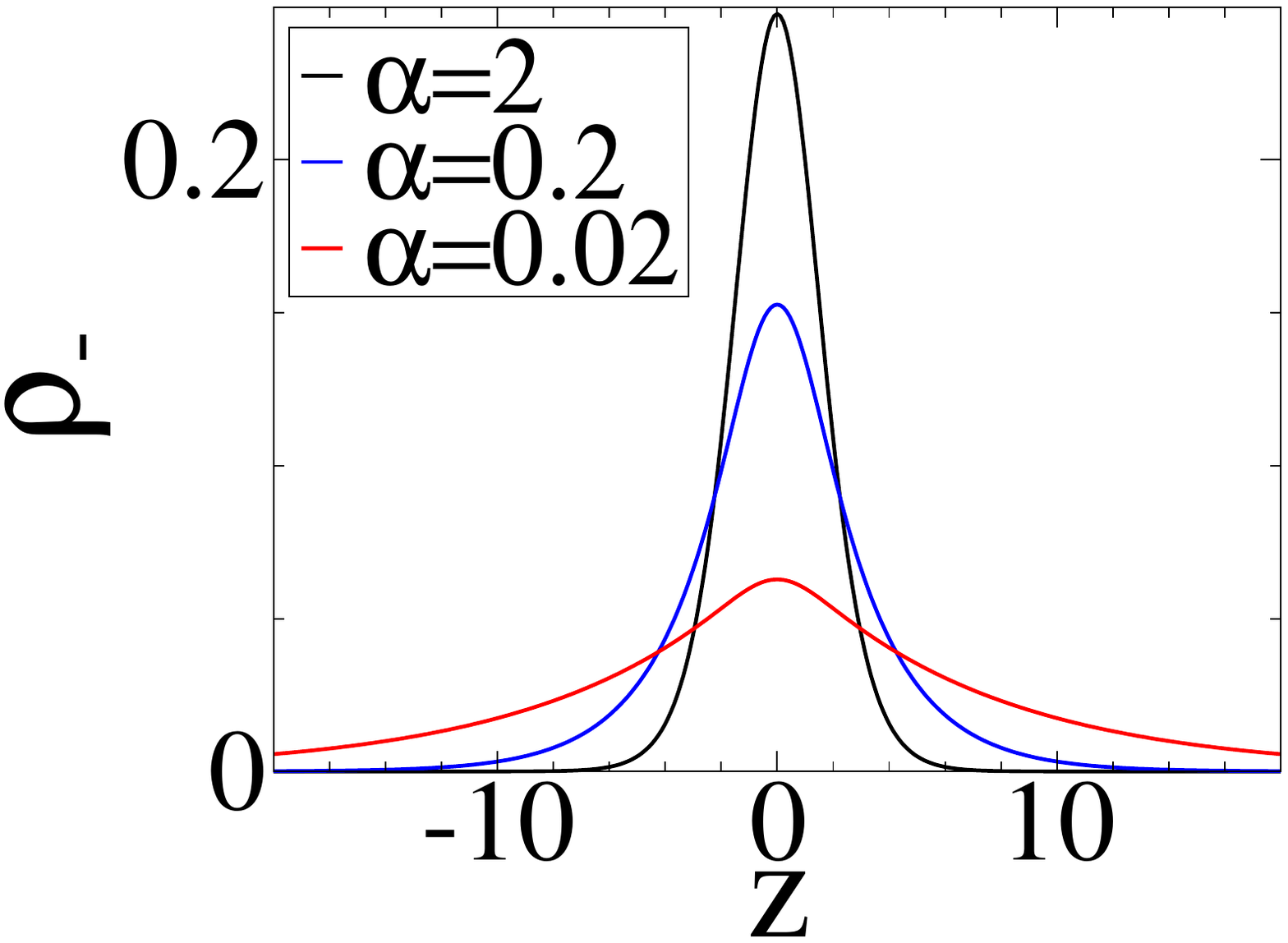} 
\end{tabular}
 \end{center} 
\caption{ Distributions of trapped and released particles, $\rho_{+}$ and $\rho_{-}$, 
obtained from the superposition formula in Eq. (\ref{eq:rho-p}) and the distributions
$p_{\pm}$ in Eq. (\ref{eq:ppm}) for different parameters $\alpha$.   } 
\label{fig:rho2} 
\end{figure}

We can get a better sense of what is happening by plotting distributions for separate states $\rho_{\pm}$.  
In Fig. (\ref{fig:rho2}) we plot distributions $\rho_{\pm}$ calculated using the superposition formula 
in Eq. (\ref{eq:rho-p}) and the distributions $p_{\pm}$ in Eq. (\ref{eq:ppm}).  
\graphicspath{{figures/}}
$\rho_+$ and $\rho_-$ respond quite differently to the reduction of 
$\alpha$.  While $\rho_+$ converges to a Gaussian distribution for the strength $\lambda=1$, 
$\rho_-$ becomes more spread out and deviates increasingly from a Gaussian form.  
Thus, the spread of particles seen in Fig. (\ref{fig:rho}) can be attributed 
to released particles.

\subsection{characteristic function}
\label{sec:sec3a}

Even though the integral in Eq. (\ref{eq:rho-full}) cannot be evaluated for an arbitrary $\alpha$, 
the characteristic function $\hat \rho(q) = \int_{-\infty}^{\infty} dz \, e^{-iqz} \rho(z)$ has a relatively 
simple algebraic form 
given by 
\be
\hat \rho(q) =  e^{-{q^2}/{2}}  \left[ \frac{ 2\alpha +q^2 }{2\alpha} \left( \frac{\alpha} {\alpha + q^2} \right)^{\frac{\alpha}{2}+1}   \right].   
\label{eq:rhoq}
\ee
Since the factor $e^{-{q^2}/{2}}$ is the characteristic function of the Boltzmann distribution, 
we can represent $\hat \rho(q)$ as 
$$
\hat \rho(q)  =  \hat \rho_{eq} (q)  \hat\rho_{neq}(q),
$$
where $\hat \rho_{eq} (q)  = e^{-{q^2}/{2}}$, and $\hat\rho_{neq}(q)$ captures non-equilibrium contributions.
As the product of two Fourier transformed function corresponds to the convolution in the real space, 
we could represent $\rho$ as  
\be
\rho(z)   =   \int dz'\,  \rho_{eq}(z-z') \rho_{neq}(z'), 
\label{eq:conv}
\ee
and because the convolution construction implies the presence of two independent random processes, 
we could interpret $\rho_{neq}$ as some type of random process, although it is not clear how to identify 
this process.   
At the crossover $\alpha=2$, $\rho_{neq}$ is found to have a simple form given by
\be
\rho_{neq} = \frac{e^{-|z| \sqrt{2} }}{8} \left( 3\sqrt{2}  +  2 z\right), 
\label{eq:rho-neq-2}
\ee
and the convolution formula in Eq. (\ref{eq:conv}) smears out this result 
into the formula in Eq. (\ref{eq:rho-full-b4}).  Despite Gaussian smearing, the asymptotic 
behavior of $\rho$ should be dominated by the exponential form of $\rho_{neq}$, which 
indicates that $\rho$ has different asymptotic behavior than that of a Boltzmann distribution.

\section{Time-dependent harmonic trap}
\label{sec:time}

In this section we provide justification for the superposition formula in Eq. (\ref{eq:rho-p}).  
The system we study is time-dependent due to constant evolution of the trap strength over time.  
However, it we average this distribution over long times, we obtain a stationary distribution.  

We start by we considering a harmonic potential with a general time-dependent strength, 
\be
u(x,t) =  \frac { K(t) x^2}{2}.  
\ee
We next assume that the time-dependent distribution of such a system
is a Gaussian function 
at all times, 
\be
\rho(x,t) =   \sqrt{ \frac{\mu K_{eff}}{2\pi D} }   \exp\left[ -\frac{\mu K_{eff} }{D} \frac{x^2}{2}  \right],
\label{eq:gauss-d}
\ee
where the only parameter that changes in time is the 
time-dependent effective strength $K_{eff}(t)$ such that $K_{eff}(t)\neq K(t)$ (unless $K(t)$ varies very slowly).  
The only constraint we introduce is that at $t=0$, $\rho(x,t)$ is a Gaussian function corresponding to 
some initial effective strength $K_0 = K_{eff}(0)$.  

To calculate $K_{eff}$, we insert the Gaussian distribution in Eq. (\ref{eq:gauss-d}) into 
the corresponding time-dependent Fokker-Planck equation, 
\be
\dot \rho   =    \mu K(t) [x\rho]'  +    D\nabla^2 \rho, 
\label{eq:diff-d}
\ee
which yields the following equation
\be
\mu K  -  \mu K_{eff}    =   \frac{1}{2} \frac{d \ln K_{eff}}{dt}, 
\label{eq:diff-Keff}
\ee
for which the solution is 
\be
K_{eff}(t)      =     \frac{ K_0   }{ e^{ -2 \int_0^{t} dt'\, \mu K(t')}    +  2   \mu K_0  \int_0^{t} dt'\, e^{ -2 \int_{t'}^{t} dt''\, \mu K(t'')}  }.  
\label{eq:keff-d}
\ee
The fact that Eq. (\ref{eq:diff-Keff}) can be solved implies that 
the Gaussian distribution in Eq. (\ref{eq:gauss-d}) is a correct time-dependent distribution and the 
solution of Eq. (\ref{eq:diff-d}).

If $K(t)$ changes in some periodic, quasi-periodic, or any other repetitive fashion, then it should be 
possible to obtain a stationary 
distribution by averaging $\rho$ over a long time, 
\be
\rho(x) = \lim_{t\to\infty}  \frac{1}{t} \int_0^{t} dt'\, \rho(x,t').
\label{eq:rhot}
\ee
And since it was determined that the distribution at all times has a Gaussian form, we could 
alternatively represent $\rho(x)$ as a superposition of Gaussian distributions, 
\be
\rho(x) =    \lim_{t\to\infty}  \frac{1}{t}  \int_0^{t} ds\, \rho_G(z,s)  =    \int_{0}^{1} d\lambda \, p(\lambda)    \rho_G(z;\lambda),
\label{eq:rho-super}
\ee
where the probability distribution $p(\lambda)$ depends on a specific evolution of $K_{eff}(t)$.  
The superposition formula  in Eq. (\ref{eq:rho-p}) is simply a consequence of this relation.  

In the case that $K(t)$ changes discontinuously as $K_0\to K_1$, Eq. (\ref{eq:keff-d}) reduces to 
\be
K_{eff}(t)      =     \frac{  K_0 K_1  }{    K_0  - (K_0   -  K_1 ) e^{ -2 \mu K_1 t}  }.  
\label{eq:keff-fast}
\ee
In the model considered in this work, the strength of the harmonic potential changes between two 
discrete values, $0$ and $K$, corresponding to the strength of the actual trap.  

\begin{figure}[hhhh] 
 \begin{center}
 \begin{tabular}{rrrr}
 \includegraphics[height=0.19\textwidth,width=0.21\textwidth]{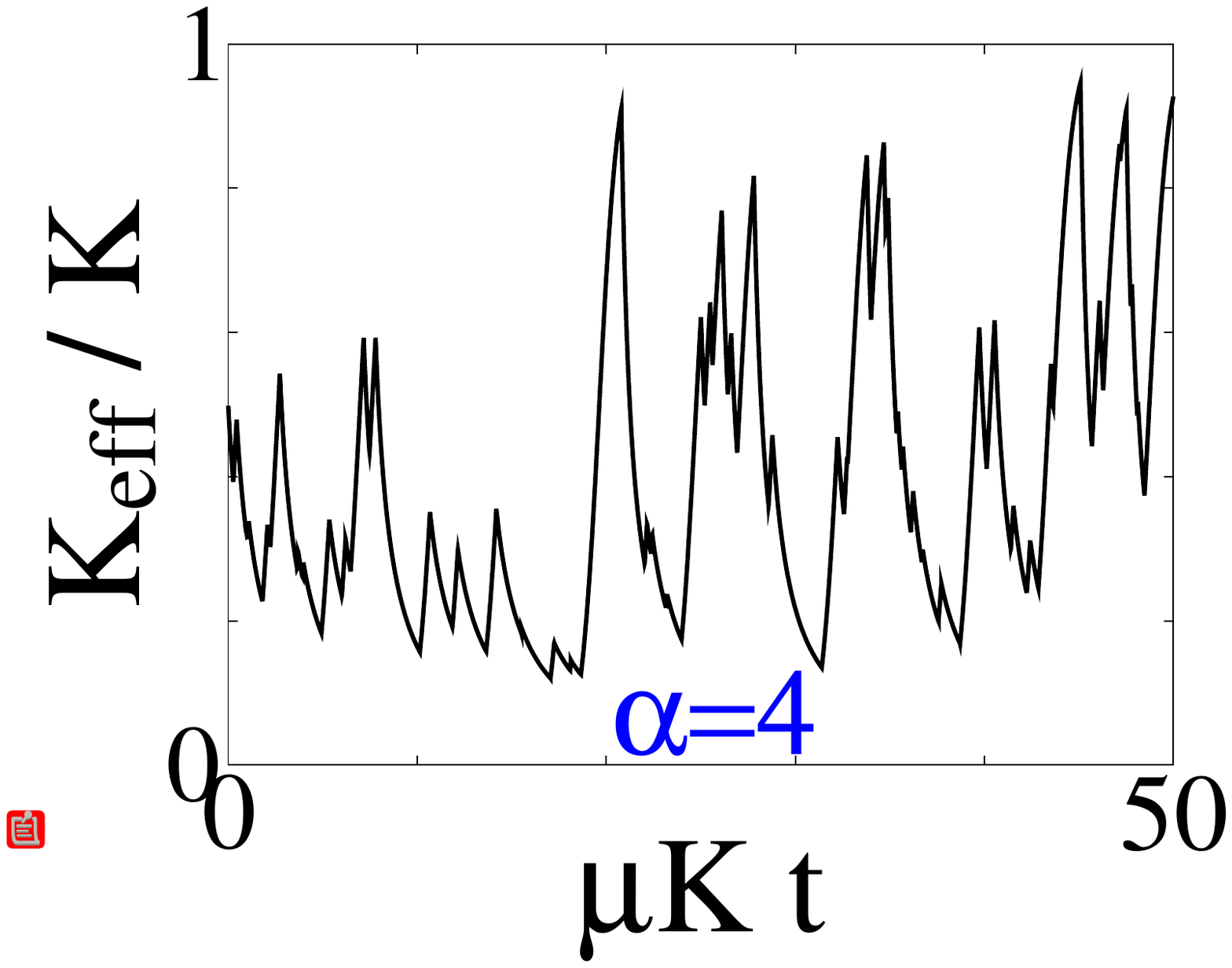} \\
 \includegraphics[height=0.19\textwidth,width=0.21\textwidth]{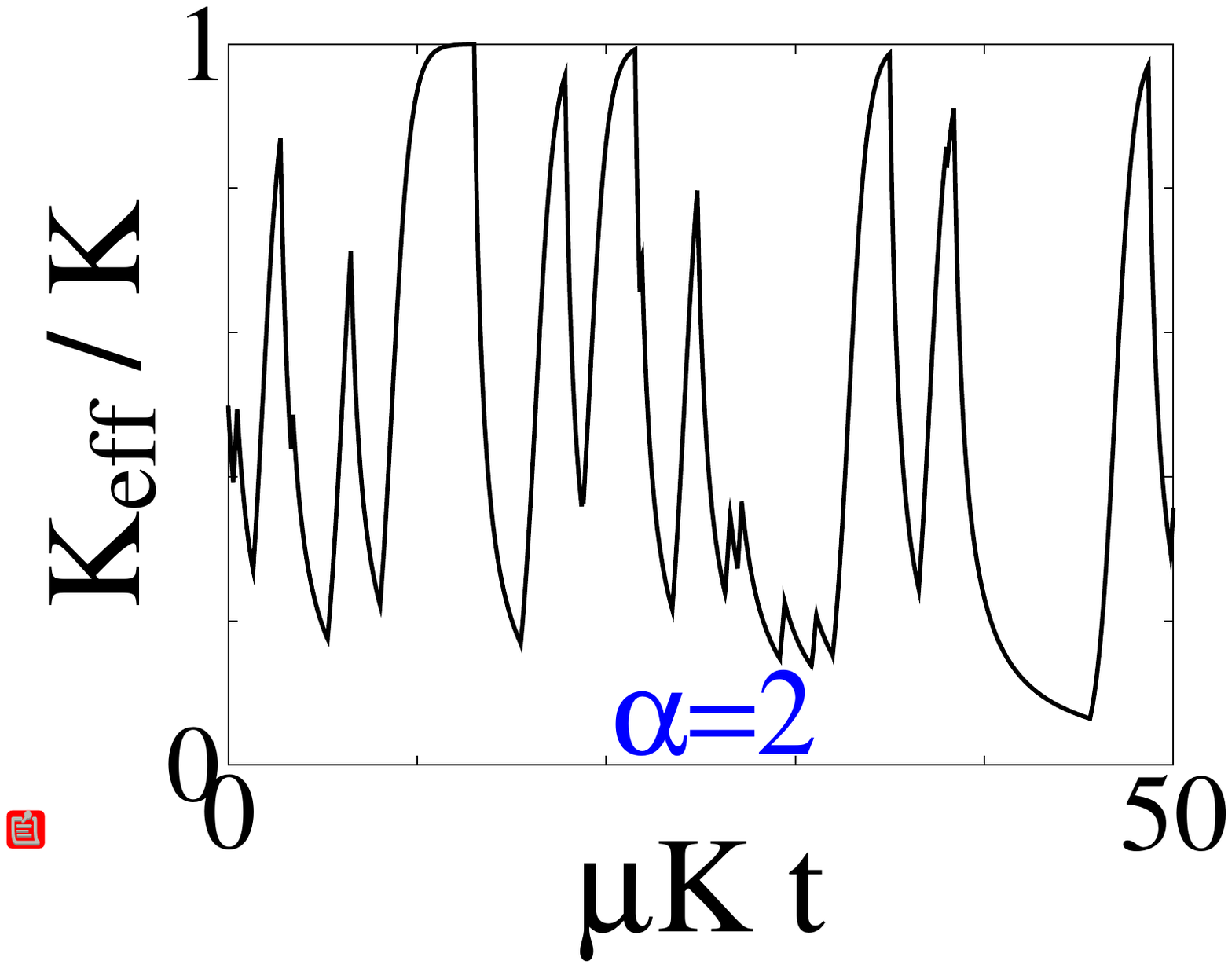} \\
 \includegraphics[height=0.19\textwidth,width=0.21\textwidth]{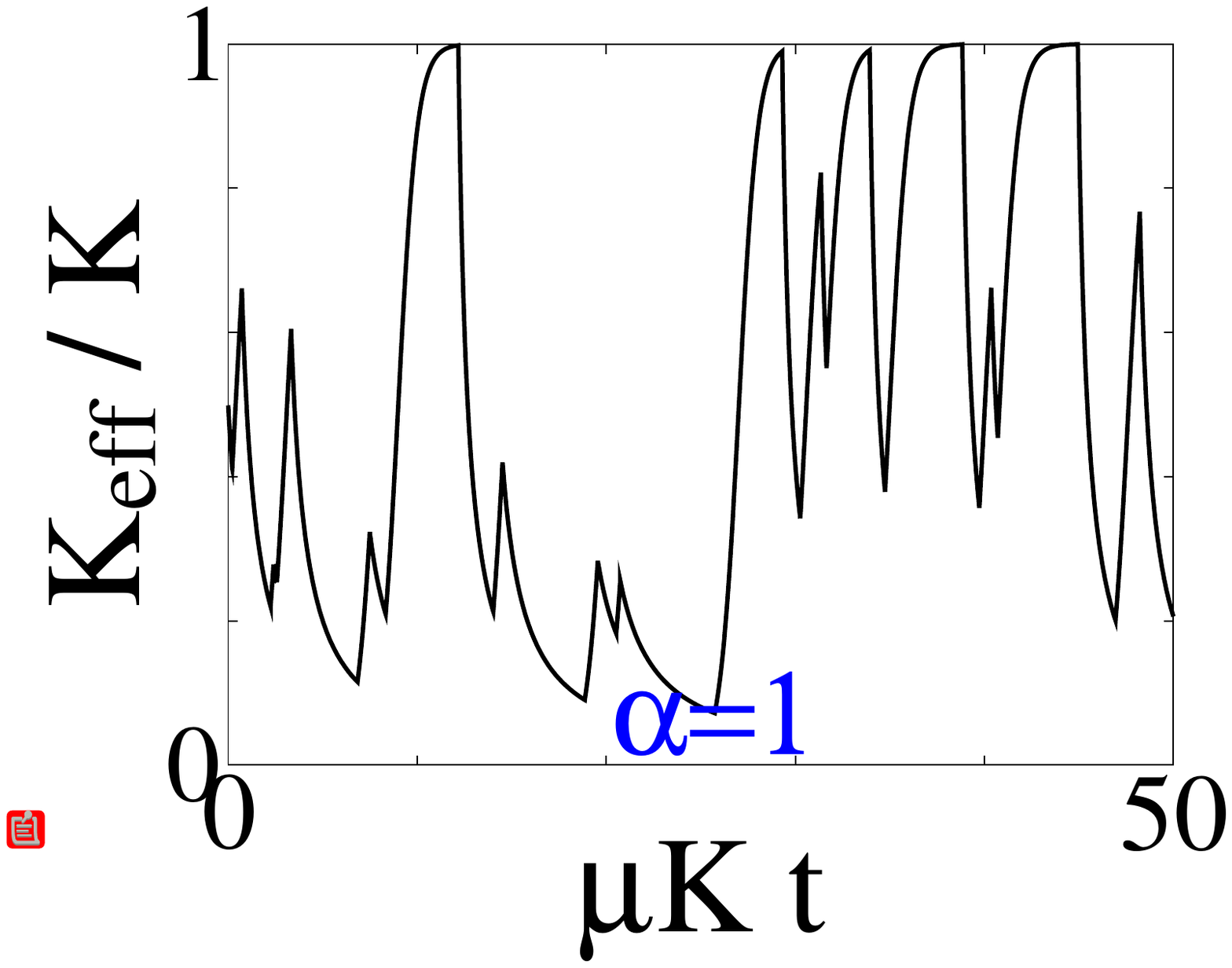} 
\end{tabular}
 \end{center} 
\caption{ $K_{eff}$ of a fluctuating trap for different $\alpha = 1/\mu K \tau$.  The initial value is 
$K_{eff}(0) = K/2$ for all figures.} 
\label{fig:keff} 
\end{figure}

Using Eq. (\ref{eq:keff-fast}), we could calculate the evolution of $K_{eff}(t)$, where $K(t)$ fluctuates 
between two values and remains in each value 
for the stretch of time drawn from the exponential distribution.  The resulting 
evolution of $K_{eff}$ is shown in Fig. (\ref{fig:keff}) for different values of $\alpha$. 
\graphicspath{{figures/}}

For $\alpha=4$, which is above the point of crossover, $K_{eff}$ fails to come close to $K$ due to rapid 
fluctuations of a trap.  This explains why the probability distribution $p(\lambda)$ in Fig. (\ref{fig:p}) for 
the same value of $\alpha$ is zero at $K_{eff}/K = \lambda=1$.  For $\alpha=1$, which is below the 
crossover, $K_{eff}$ can approach $K$ and then remain close to it for extended periods of time.  
This gives rise to a divergence in 
$p(\lambda)$ at $\lambda = 1$ seen in Fig. (\ref{fig:p}) for the same value of $\alpha$.  

One can obtain the distribution $p(\lambda)$ from the evolution of $K_{eff}(t)$ using 
$p(\lambda) \sim \lim_{t\to\infty} \frac{1}{t} \frac{dt}{d\lambda}$, where $\lambda(t) = K_{eff}(t)/K$.

\section{Extension to higher dimensions}
\label{sec:extension}

It is rather straightforward to extend the results in Sec. (\ref{sec:time}) to an arbitrary dimension $d$.  
Given a time-dependent harmonic potential, 
\be
u(r,t) =  \frac { K(t) r^2}{2}, 
\ee
where $r^2 = \sum_{i=1}^d x_i^2$ is the radial distance from a trap center for a system 
in dimension $d$.  
We next assume that the time-dependent distribution has a Gaussian form at all times, 
\be
\rho(r,t) =   \left( \frac{\mu K_{eff}}{2\pi D} \right)^{d/2}  \exp\left[ -\frac{\mu K_{eff} }{D} \frac{r^2}{2}  \right].  
\label{eq:gauss-dd}
\ee
To obtain an expression for $K_{eff}(t)$, we insert the Gaussian distribution 
above into the corresponding Fokker-Planck equation, 
\be
\dot \rho   =    \mu K(t) \bnabla\cdot ( {\bf r} \rho )   +    D\nabla^2 \rho.  
\label{eq:diff-ds}
\ee
Such a procedure recovers the relation in Eq. (\ref{eq:diff-Keff}) for which the solution is the expression 
in Eq. (\ref{eq:keff-d}).  This proves that $K_{eff}(t)$ in Eq. (\ref{eq:keff-d}) is valid for any dimension.    
Other results in this work can also be extended to an arbitrary dimension.  
The Fokker-Planck equation for the fluctuating potential model, analogous to Eq. (\ref{eq:FP}) but for an 
arbitrary dimension, is given by 
\ba
&&\dot\rho_+ =   \mu K\bnabla \cdot [{\bf r} \rho_+]  +  D\nabla^2 \rho_+  +  \frac{1}{\tau}  \left(\rho_-  -  \rho_+\right) \nonumber\\ 
&&\dot\rho_-  =    D\nabla^2 \rho_-   -    \frac{1}{\tau}   \left(\rho_-  -  \rho_+\right).  
\ea
For a stationary state and in reduced units the two equations become 
\ba
&& 0   =    \bnabla \cdot [{\bf s} \rho_+]  +  \nabla^2 \rho_+  +  \alpha \left(\rho_-  -  \rho_+\right) \nonumber\\ 
&& 0   =    \nabla^2 \rho_-  -  \alpha \left(\rho_-  -  \rho_+\right), 
\label{eq:FPS-d}
\ea
where ${\bf s} = {\bf r} \sqrt{\frac{\mu K}{D}}$.

The moments of stationary distributions can be obtained by operating on both equations in 
Eq. (\ref{eq:FPS-d}) with $\int d{\bf r}\, r^{d-1} \, r^{2n}$.  
The resulting expressions for the second moment are  
\ba
&&  \langle s^2 \rangle_{+}     =     2 d         \nonumber\\  
&&  \langle s^2 \rangle_{-}      =     2 d   \left( 1   +   \frac{1}{\alpha} \right)         \nonumber\\     
&&  \langle s^2 \rangle            =      d   \left( 2   +   \frac{1}{\alpha} \right).     
\label{eq:s2}
\ea


By combining the two equations in Eq. (\ref{eq:FPS-d}) we obtain a third-order differential equation, 
\be
0     =     -\alpha s^3 \rho      +     [3 - d - (2\alpha + 1 - d) s^2 ]   \rho'        -      [3 - d - s^2] s \rho''      +      s^2 \rho'''.  
\label{eq:diff-3-d}
\ee
Note that by setting $d=1$ we recover Eq. (\ref{eq:diff-3}).  
To find the solution of the above equation, we represent a stationary distribution as a superposition of Gaussian distributions,
\be
\rho(s)   =   \int_{0}^{1} d\lambda \, p(\lambda)    \rho_G(s;\lambda), 
\label{eq:rho-p-d}
\ee
where 
\be
\rho_G(s;\lambda) = \left( \frac{\lambda}{2\pi}\right )^{d/2} e^{-\lambda s^2/2}.  
\label{eq:rho-G-d}
\ee


Since it was determined that $K_{eff}(t)$ is independent of $d$, and since it is possible to obtain the 
probability distribution $p(\lambda)$ from the evolution of $K_{eff}(t)$, we conclude that the expression for 
$p(\lambda)$ in Eq. (\ref{eq:p-lambda}) is valid for all $d$.  This means that any 
dependence on $d$ comes from the Gaussian distribution in Eq. (\ref{eq:rho-G-d}), 
and the solution of the third-order differential equation in Eq. (\ref{eq:diff-3-d}) is given by 
\be
\rho(s)   =   \frac{ ( \alpha e /2)^{\alpha/2} }{ 2\Gamma(\alpha/2) }  \left(\frac{1}{2\pi } \right)^{\frac{d}{2}}   \int_0^1 d\lambda\,   
\left( \frac{1}{\lambda} - 1\right)^{\frac{\alpha}{2} - 1}  \frac{e^{-\frac{\alpha}{2\lambda}} e^{-\frac{\lambda  s^2}{2}} } { \lambda^{3-\frac{d}{2}} }.
\label{eq:rho-full-d}
\ee
We note that the solution is not limited to integer values of $d$ and applies to any value of $d$.

\section{Quantities of physical interest}
\label{sec:quantities}

\subsection{potential energy}


In this section we look into physical quantities of the model, starting with the average potential energy 
$\langle u\rangle_{\pm} = K\langle r^2\rangle_{\pm}/2$.  
Using the moments in Eq. (\ref{eq:s2}), the potential energy for particles 
in each state is found to be
\ba
&& \langle u \rangle_+  =  d k_BT, \nonumber\\
&& \langle u \rangle_-  =  0.  
\ea
The potential energy of particles in a released state is zero since the trap 
is switched off.  Another interesting observation is that the potential energy of particles in a trapped
state does not depend on the rate parameter $\alpha$.  
Adding the two contributions, the total potential energy is found to be the same as that for a system in equilibrium, 
\be
\langle u \rangle =  \frac{d k_BT}{2}.  
\label{eq:PE}
\ee
The fluctuating harmonic potential might considerably alter stationary distributions but the average potential 
energy is unaffected.  This is a bit unexpected, however, at least the mathematics of this result 
can be traced to the fact that the second moment
of trapped particles, see Eq. (\ref{eq:s2}), does not depend on $\alpha$.

\subsection{flux}

We next consider flux, which for particles in each state and for $d=1$ is given by  
\be
j_-   =  -D\rho'_- ,   ~~~~~ j_+   = - \mu K  x \rho_+  -   D\rho'_+. 
\ee
Also note that the total flux is zero everywhere, $j = j_+  +  j_- = 0$. 

\begin{figure}[hhhh] 
 \begin{center}
 \begin{tabular}{rrrr}
 \includegraphics[height=0.19\textwidth,width=0.21\textwidth]{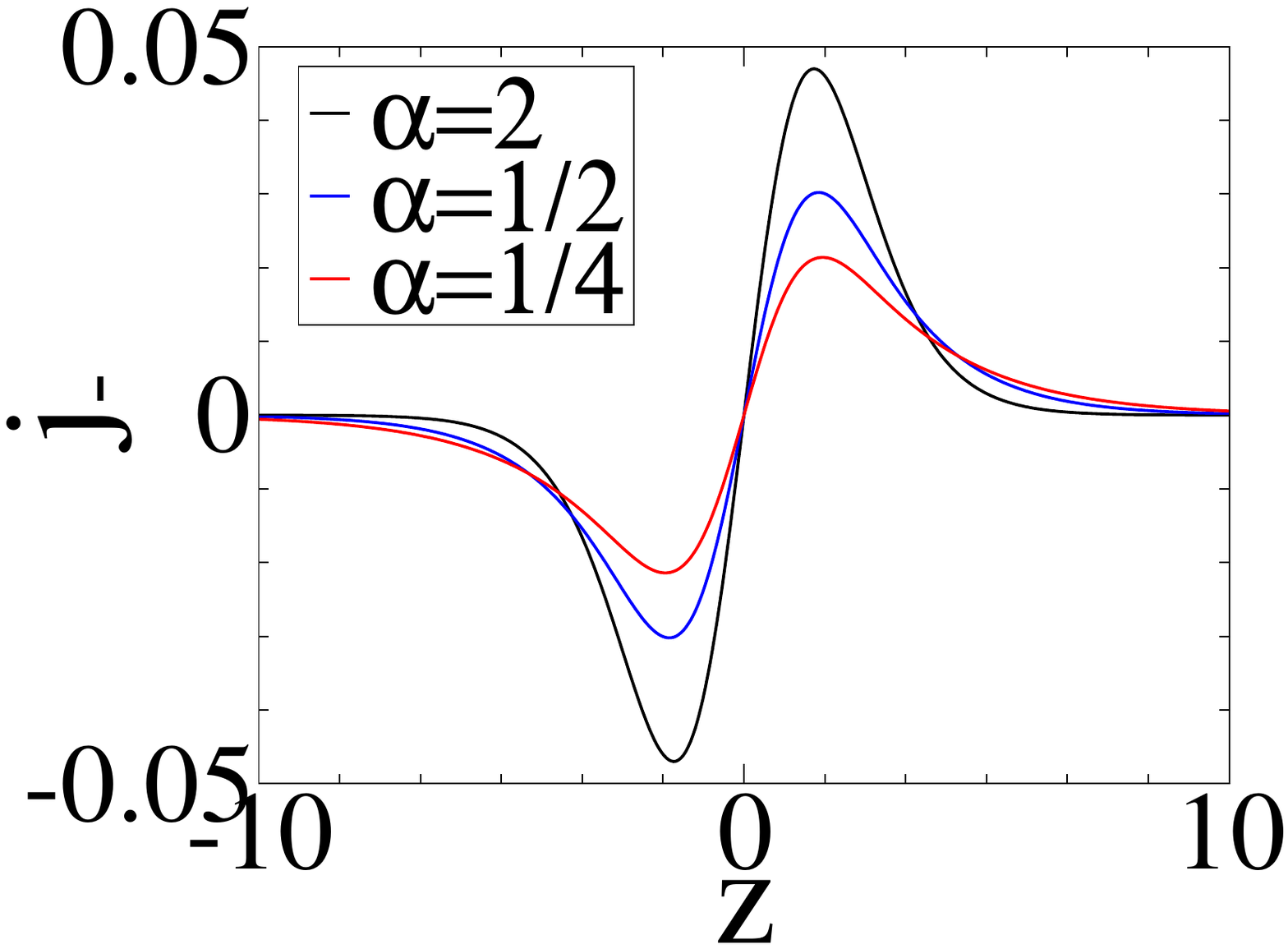} &
\end{tabular}
 \end{center} 
\caption{ Flux for particles released from the trap.   
Flux for particles in a trapped state, not shown in the figure, is $j_+ = - j_-$.  } 
\label{fig:ja} 
\end{figure}

In Fig. (\ref{fig:ja}) we plot flux for particles in a released state, $j_-$, in dimensionless units
given by $j_-(z) = -\rho'_-(z)$.  (No insight is gained by plotting $j_+$ since $j_+ = -j_-$).
The first observation is that released particles move away from the trap center, which is 
expected because particles are released.  The second observation is that the 
magnitude of the flux increases with increased $\alpha$.  This makes sense since released 
particles tend to be more compressed when $\alpha$ is large and so diffusion away from the 
trap center should be greater.   
\graphicspath{{figures/}}

As both fluxes have the same functional form, a better insight could be gained by looking into the local velocity, 
in dimensionless units given by 
\be
v_+   =   \frac{j_+}{\rho_+}     =    -[\ln \rho_+]'  -  z, 
~~~~~~ v_-    =    \frac{j_-}{\rho_-}   =   -[\ln \rho_-]'. 
\label{eq:vpm}
\ee
Note that the velocity in a trapped state has an additional linear term resulting 
from an external force.  The other term in each expression comes from diffusion.

\begin{figure}[hhhh] 
 \begin{center}
 \begin{tabular}{rrrr}
 \includegraphics[height=0.19\textwidth,width=0.21\textwidth]{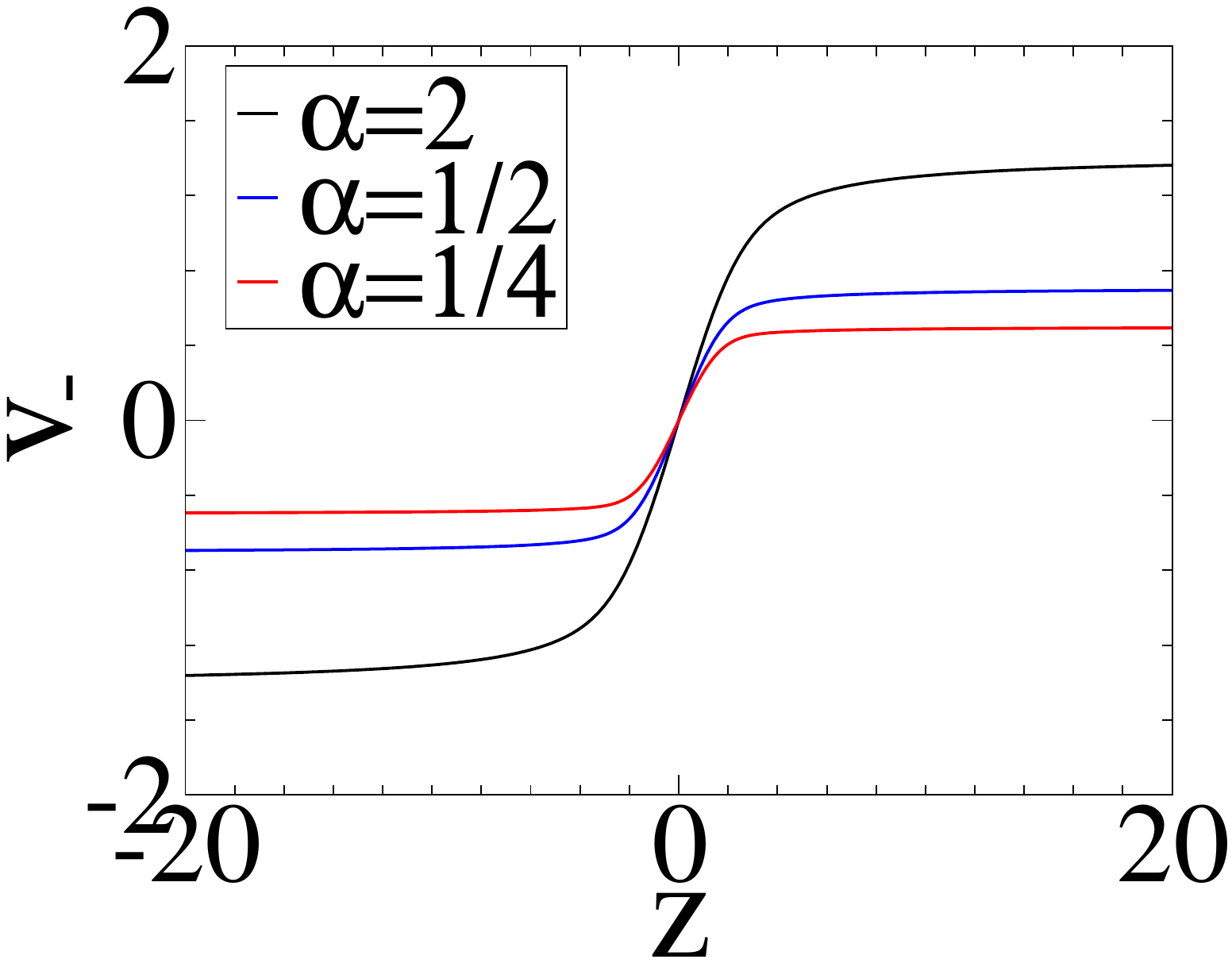} &
 \includegraphics[height=0.19\textwidth,width=0.21\textwidth]{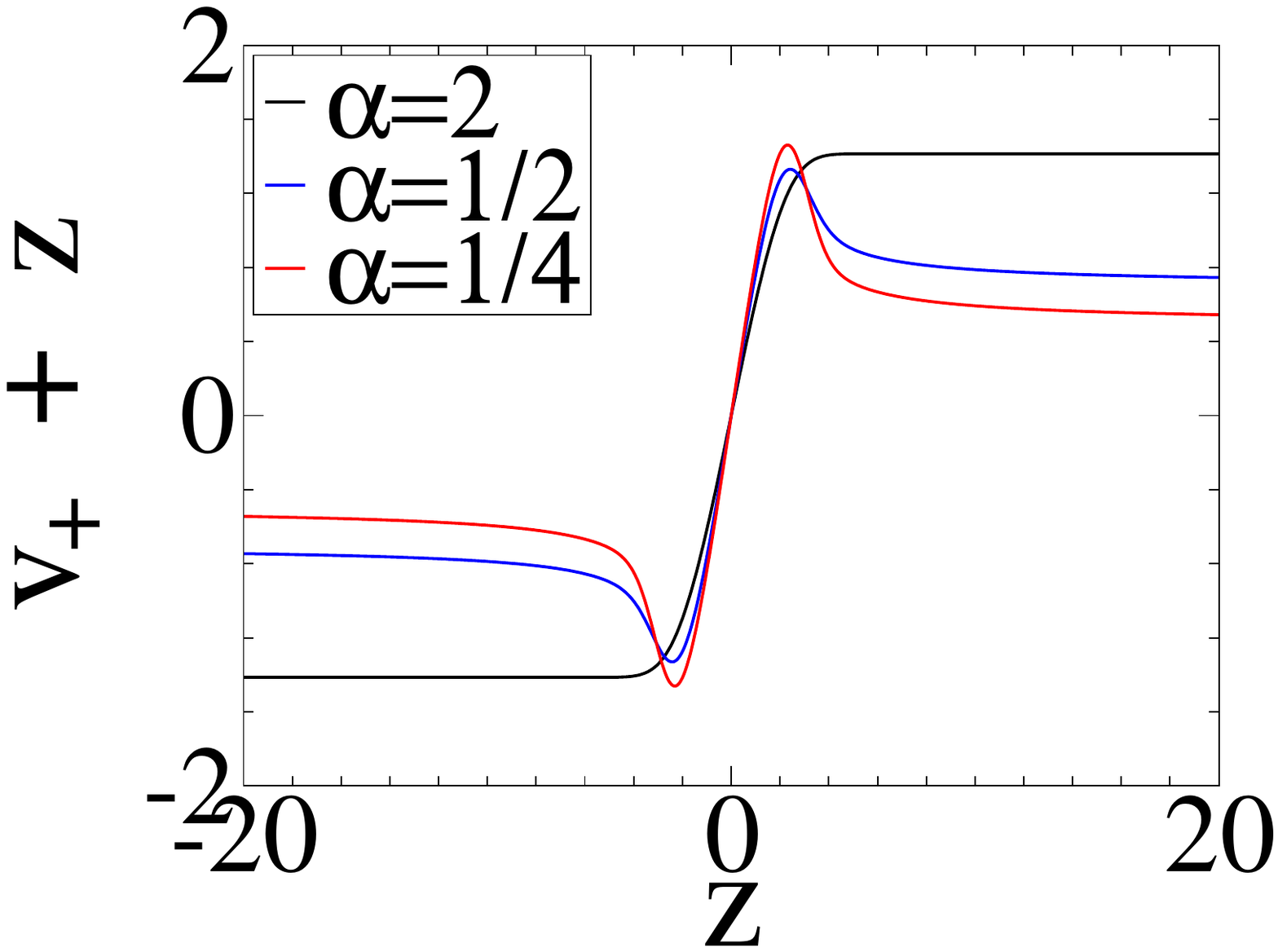} &
\end{tabular}
 \end{center} 
\caption{Local velocity for different values of $\alpha$.  For trapped particles we plot $v_+ + z$ 
to subtract contributions of an external force, as a result, both quantities represent diffusional 
component of a local velocity.  In both cases this corresponds to plotting $-[\ln\rho_{\pm}]'$.} 
\label{fig:va} 
\end{figure}

Local velocities are shown in Fig. (\ref{fig:va}).  For 
particles in a trapped state we plot $v_+ + z$ to subtract contributions of an external 
force.  Consequently, the two plots amount to $-[\ln \rho_{\pm}]'$ and represent 
diffusional component of a local velocity.  
\graphicspath{{figures/}}

The most striking feature is that the local velocity does not vanish at infinity but 
saturate at a non-zero constant.  Since the quantity that is plotted is $-[\ln\rho_{\pm}]'$, this implies  
that distributions $\rho_{\pm}$ decay exponentially.  Such an asymptotic behavior has been hinted 
at in Sec. (\ref{sec:sec3a}) and was proven for $\alpha=2$ in Eq. (\ref{eq:rho-neq-2}).  

\subsection{entropy production rate}
\label{sec:heat}

One feature of an active system is that it generates heat that is dissipated into reservoir. 
In the case of active particles, it is the internal energy of a particle that is converted to heat.
In the case of a present system, the heat is generated when the work done by an external 
potential to compress particles is converted to heat.   

By measuring the rate of heat that is dissipated, we could quantify a distance of how far 
from equilibrium a system is, since the rate of heat dissipation, $\langle \dot q\rangle$, is 
related to the entropy production rate $\Pi$ via the relation $T\Pi \equiv \langle \dot q\rangle$, 
where $T$ is the temperature \cite{Frydel23a}.

There are different ways one can obtain an expression for the entropy production rate \cite{Cocconi22}.  
Here we are going to follow an intuitive approach.   A particle that is moving through a dissipating medium 
experiences the drag force ${\bf F}_d = -\mu^{-1} {\bf v}$, where ${\bf v}$ is the velocity vector.  
An instantaneous rate of heat dissipation is then given by $\dot q = \mu^{-1} v^2$.  However, when 
calculating the average rate of heat dissipation, we need to subtract the equilibrium contributions, 
$\langle \dot q\rangle = \mu^{-1} \langle v^2\rangle - \mu^{-1} \langle v^2\rangle_{eq}$, since these 
contributions are not truly dissipated but are recovered at some point in the form of thermal 
fluctuations as a result of the fluctuation-dissipation relation.  And since $\langle v^2\rangle_{eq} = dk_BT/m$, 
we can write 
\be
\langle \dot q\rangle = \frac{1}{\mu} \left[ \langle v^2\rangle - \frac{d D}{\tau_r} \right],
\label{eq:q}
\ee
where $\tau_r=m\mu$ is the inertial relaxation and $m$ is the mass of a particle.  For overdamped
dynamics there is no inertia and $\tau_r=0$.

To calculate the average velocity $\langle v^2\rangle$ we formulate our model in the underdamped regime. 
The formulation and subsequent derivations are carried out in Appendix \ref{app:sec4}.  Here we write down the 
end result, 
\be
\langle v^2 \rangle - \frac{d D}{\tau_r} =  \frac{d  \mu K D }{2} \frac{  (1 + \tau_r/\tau )} {  1  +   3 \tau_r/ \tau    +     \tau_r^2/\tau^2 (2   -  \mu K \tau/2) },
\label{eq:v2-taur-T}
\ee
and remind that $\tau$ is the average time during which a system remains in a given state.  
Inserting the above expression in Eq. (\ref{eq:q}) yields 
\be
T\Pi \equiv   \langle \dot q\rangle   =      \frac{d K D }{2} \frac{  (1 + \tau_r/\tau )} {  1  +   3 \tau_r/ \tau    +     \tau_r^2/\tau^2 (2   -  \mu K \tau/2) }. 
\label{eq:EP-taur}
\ee
For the overdamped regime $\tau_r=0$, the expression above reduces to 
\be
T\Pi \equiv   \langle \dot q\rangle   =    \frac{d D K}{2}.  
\label{eq:EP}
\ee
The entropy production rate for a fluctuating harmonic trap in the overdamped regime 
has been previously obtained in \cite{Cocconi22}.  Our Eq. (\ref{eq:EP}) 
is in agreement 
with that result.  


The main difference between the result in Eq. (\ref{eq:EP}) and the dissipation of heat of typical systems of 
active particles in a harmonic trap, such as 
run-and-tumble and active Brownian particles, is that $ \langle \dot q\rangle$ in these systems 
depends on $\alpha$ as $\langle \dot q\rangle \propto \alpha/(1+\alpha)$ \cite{Frydel23a}, where $\alpha$ 
represents the rate at which an active particle changes direction of motion.  According to this relation, 
$\langle \dot q\rangle$ is maximal in the limit $\alpha\to \infty$ and 
then decreases monotonically to zero as $\alpha\to 0$.   The reason that $\langle \dot q\rangle$ vanishes 
as $\alpha\to 0$ is that the direction of motion in this limit changes very seldom and there is sufficient time 
for a system to attain equilibrium.  In the current system, the attainment of equilibrium even in the limit $\alpha\to 0$
is not possible --- equilibrium for particles in a released state corresponds to particles that are 
spread out into infinity.  Attaining this distribution would require an infinite amount of time.  As a result, 
particles in a harmonic trap remain agitated even in the limit $\alpha\to \infty$.  

We next look into the formula for the dissipation of heat in the underdamped regime $\tau_r>0$ in Eq. (\ref{eq:EP-taur}).  
The first observation is that $\langle \dot q\rangle$ is no longer independent of $\tau$ (or $\alpha = 1/\tau \mu K$).  
Even if inertia might be negligible for active systems in a liquid, and the formula in Eq. (\ref{eq:EP}) accurately 
predicts the dissipation of heat for a significant range of $\tau$, 
the dependence on $\tau$ becomes 
important in the limit $\tau\to 0$ (or $\alpha\to \infty$) where $\langle \dot q\rangle$ is found to vanish.  
This is a drastically different limiting behavior than that deduced from the overdamped regime.  
Since the vanishing dissipation of heat indicates equilibrium, 
it means that the system in the limit $\alpha\to \infty$ is in equilibrium, however, we only reach this conclusion 
when we incorporate inertia
into a theoretical description.  Since underdapmed dynamics represents a more complete description, we conclude
that the limit $\alpha\to \infty$ indeed represents a true equilibrium.  
This claim is supported by the fact that a 
stationary distribution in this limit converges to a Gaussian distribution, see Eq. (\ref{eq:diff-3-lim}).    


\section{Conclusion}
\label{sec:conclusion}

This work considers an ideal gas confined to a harmonic trap $u=Kr^2/2$ with  
time dependent strength $K\equiv K(t)$.  It is demonstrated that the time dependent distribution of 
such a system $\rho(x,t)$ has a Gaussian form at all times, corresponding to some effective
strength of the harmonic potential, $K_{eff}(t)\neq K(t)$.   

We are interested in a specific time evolution of $K(t)$ such that it changes discontinuously
between two discrete values, $0$ and $K$, and where the time during which a trap remains
in a given state is drawn from an exponential distribution $\sim e^{-t_p/\tau}$.


A stationary distribution of this system can be obtained from a third-order differential 
equation, the solution of which is represented as a superposition of Gaussian 
distributions for different strengths of a harmonic trap, $\rho(x) = \int_0^{K} d K' \, p(K') \rho_G(x;K')$.  
The reason for the superposition can be traced back to the fact that a time-dependent distribution
at all times has a Gaussian form.  

The probability distribution $p(K)$ can be obtained and analyzed.  
The resulting algebraic expression for $p(K)$ exhibits a crossover at $\tau = 2/\mu K$.  For $\tau > 2\mu K$, 
the distribution $p(K')$ diverges at $K'=K$, indicating that a fraction of particles comes close to an 
equilibrium-like behavior.  The divergence disappears for $\tau < 2/\mu K$.  
For very small $\tau$, a stationary distribution converges to a Gaussian form for a trap with the 
strength which is half the strength of the physical trap, $K/2$.

The primary feature of a resulting stationary distribution $\rho(x) = \int_0^{K} d K' \, p(K') \rho_G(x;K')$  
is the spread-out caused by the periods during which a trap is turned off.  Since the period 
during which a trap is turned off increases with $\tau$, the distributions tend to be more spread out 
for larger $\tau$.  The spreading component of distributions is dominated
by an exponential decay.  

The spreading-out feature of the current model is very different from the  
behavior of active particles.  Active particles are known to accumulate near the trap boundaries 
rather than penetrate these boundaries \cite{Brady16,Orlandini18}.  
For example, in the case of active particles trapped in a harmonic trap, the accumulation of particles 
at trap borders leads to bimodal stationary distributions, with excess of particles close to boundaries 
and depletion of particles near a trap center, \cite{Tailleur09,Basu20,Frydel22c}.  
These models exhibit crossover when a stationary 
distribution changes from a bimodal to unimodal distribution.


\begin{acknowledgments}
D.F. acknowledges financial support from FONDECYT through grant number 1241694.  
\end{acknowledgments}

\section{DATA AVAILABILITY}
The data that support the findings of this study are available from the corresponding author upon 
reasonable request.

\appendix

\section{Derivation of Eq. (\ref{eq:diff-3})}
\label{app:sec1}

In this section we show how the two coupled differential equations in Eq. (\ref{eq:FPS}), which we reproduce below,
\ba
&&   0     =        [z\rho_+]'  +    \rho_+''      +      \alpha    \left( \rho_-   -  \rho_+ \right),   \nonumber\\
&&   0      =      \rho_-''      +      \alpha   \left( \rho_+  -  \rho_- \right),
\label{eq:app1a}
\ea
are combined to yield a single differential equation in Eq. (\ref{eq:diff-3}).  
The two equations can be either added or subtracted, where each procedure yields 
\ba
&&   0     =      \frac{1}{2} [z\rho]'   +  \frac{1}{2}  [z\sigma]'  +    \rho''    ,   \nonumber\\
&&   0      =       \frac{1}{2}  [z\rho]'   +  \frac{1}{2} [z\sigma]'    +  \sigma''      -     2\alpha \sigma,
\label{eq:app1b}
\ea
where $\rho  =  (\rho_+  +  \rho_-)/2$ and $\sigma   =   (\rho_+   -   \rho_-)/2$.  
From the first equation we get
$
z \sigma      =     - z\rho   -    2\rho'  +  \text{const},
$
where $\text{const} = 0$ as a result of the even symmetry of $\rho(z)$.  
From this we can generate a sequence of expressions 
\ba
&& \sigma      =     -\rho   -     \frac{2}{z} \rho',  \nonumber\\
&& \sigma'      =     -\rho'      +     \frac{2}{z^2} \rho'  -   \frac{2}{z} \rho'',  \nonumber\\
&& \sigma''      =     -\rho''     -    \frac{4}{z^3} \rho'   +    \frac{4}{z^2} \rho''     -    \frac{2}{z} \rho'''.
\ea
Inserting these expressions into the second equation in Eq. (\ref{eq:app1b}) recovers Eq. (\ref{eq:diff-3}).

%

\section{Recurrence relation}
\label{app:sec2a}

In this section we show how to obtain even moments, defined as  
$$
\langle z^{2n} \rangle_{\pm}  =  \int_{-\infty}^{\infty} dz\, z^{2n} \rho_{\pm}(z), 
$$
from the two equations in Eq. (\ref{eq:FPS}), which we repeat below for clarity:  
\ba
&&   0     =        [z\rho_+]'  +    \rho_+''      +      \alpha    \left( \rho_-   -  \rho_+ \right)    \nonumber\\
&&   0      =      \rho_-''      +      \alpha   \left( \rho_+  -  \rho_- \right).  
\ea
To convert the two equations into the recurrence relation, we operate on both equation with $\int_{-\infty}^{\infty} dz\, z^{2n+2}$.  
This results in the following two coupled recurrence relations:
\ba
&&   0     =        -(2n+2)m_{n+1}^+  +   (2n+2)(2n+1)m_{n}^+       +      \alpha    \left(m_{n+1}^-    -  m_{n+1}^+  \right)    \nonumber\\
&&   0      =       (2n+2)(2n+1)m_{n}^-      +      \alpha   \left( m_{n+1}^+   -  m_{n+1}^- \right),
\ea
where $m_{n}^+ = \langle z^{2n} \rangle_+$ and $m_{n}^- = \langle z^{2n} \rangle_-$.  
After rearrangement, the two equations become
\ba
&&    m_{n+1}^+     =     (2n+1) (m_{n}^+ + m_{n}^- )           \nonumber\\
&&    m_{n+1}^+     =     (2n+1) \left( m_{n}^+   -  m_{n}^- \right)    -    \frac{2\alpha}{(2n+2)} \left( m_{n+1}^+   -  m_{n+1}^- \right).  \nonumber
\ea
The initial terms of both sequences are $m_0^{+} = m_0^{-} = 1$.  The remaining terms can 
now be obtained from the recurrence relations above.

\section{Derivation of Eq. (\ref{eq:FP-p})}
\label{app:sec2}

In this section we demonstrate how combining the two equations in Eq. (\ref{eq:FP-p}), which we reproduce below, 
\ba
&& 0 =         [\lambda (\lambda  -  1)  p_+]'     +    \frac{\alpha}{2}   ( p_-  -  p_+ )   \nonumber\\
&& 0 =         [\lambda^2 p_-]'        +         \frac{\alpha} {2} ( p_+  -  p_- ), 
\label{eq:app2a}
\ea
yields Eq. (\ref{eq:ppm}).
Adding and subtracting the two equations yields  
\ba
&& 0 =          [\lambda (2\lambda-1)  p]'     -    \lambda \sigma'         \nonumber\\
&& 0 =        2 [\lambda^2 \sigma]'         -  \lambda p'    -    \lambda \sigma'     -     2\alpha \sigma, 
\label{eq:app2b}
\ea
where $p  =  (p_+   +   p_-)/2$ and $\sigma   =   (p_+   -   p_-)/2$.    
From the first equation we get
$
\sigma   =   (2\lambda-1)  p  
$
and from which we generate the consecutive expressions 
\ba
&& \sigma   =   (2\lambda-1)  p,  \nonumber\\
&&  \sigma'   =  2 p  +   (2\lambda-1)  p', 
\ea
Inserting these expressions into the second equation in Eq. (\ref{eq:app2b}) recovers Eq. (\ref{eq:ppm}).

\section{ Derivation of $\langle v^2 \rangle$}
\label{app:sec4}

In Sec. (\ref{sec:heat}) in Eq. (\ref{eq:q}) we provide an expression for the dissipation of heat
which includes the quantity $\langle v^2 \rangle$.  To calculate $\langle v^2 \rangle$
we formulate the problem within underdamped dynamics, which will take us slightly beyond 
the objectives of the present work.  

To begin with, we formulate the problem within the Kramer's equation, which is 
the type of a Fokker-Planck equation but that includes inertia \cite{Frydel23a},   
\ba
\frac{\partial\rho_+}{\partial t}  &=& -{\bf v} \cdot \bnabla_{r}\rho_+ 
+ \frac{1}{\tau_r} \bnabla_v \cdot \left[ \left(  \mu K {\bf r}  +  {\bf v}  \right) \rho_+ \right]  + \frac{D}{\tau_r^2}  \nabla_v^2 \rho_+ \nonumber\\
&+& \frac{ 1}{\tau} (\rho_-   -   \rho_+)
\nonumber
\ea
\ba
\frac{\partial\rho_-}{\partial t}  &=& -{\bf v} \cdot \bnabla_{r}\rho_- 
+ \frac{1}{\tau_r} \bnabla_v \cdot \left[ {\bf v}  \rho_- \right]  + \frac{D}{\tau_r^2}  \nabla_v^2 \rho_- 
+ \frac{1}{\tau} ( \rho_+   -   \rho_-), \nonumber
\ea
where $\rho_{\pm}\equiv \rho_{\pm}({\bf r},{\bf v},t)$, and we recall that $\tau_r = m\mu$ is the inertial 
relaxation, $m$ is a particle mass, and the overdamped regime is recovered for $\tau_r=0$.  


To calculate $\langle v^2\rangle$, or other average quantities of interest, we 
operate on the stationary Kramer's equation, $\dot \rho = 0$, with an integral operator 
$
\hat O_g = \int d{\bf v}\int d{\bf r}  \, \, g({\bf r},{\bf v}),
$
where the function $g$ is going to be defined later.  Using integration by parts, the terms of 
a transformed Kramer's equation can be represented as average quantities, 
designated by the angular brackets $\langle \dots \rangle = \int d{\bf v}\int d{\bf r}  \, \, (\dots)$.  
The resulting two equations are 
\ba
0 &=& \langle {\bf v}\cdot \bnabla_r g\rangle_+   -
\frac{1}{\tau_r}  \langle  \left(  \mu K {\bf r}  +  {\bf v}  \right)  \cdot  \bnabla_v g \rangle_+ 
+ \frac{D}{\tau_r^2}  \langle  \nabla_v^2 g \rangle_+  \nonumber\\ 
&+& \frac{\langle g \rangle_-}{\tau}   - \frac{ \langle g \rangle_+ }{\tau}
\nonumber
\ea
\ba
&& 0 = \langle {\bf v}\cdot \bnabla_r g\rangle_-   -
\frac{1}{\tau_r}  \langle   {\bf v}  \cdot  \bnabla_v g \rangle_- 
+ \frac{D}{\tau_r^2}  \langle  \nabla_v^2 g \rangle_-  
+ \frac{\langle g \rangle_+ }{\tau}   -   \frac{\langle g \rangle_-}{\tau}.  \nonumber
\ea
%
%
%
%
%
Using $g = v^2$, $g=r^2$, and $g= {\bf r}\cdot {\bf v}$, we generate from the relation above 
six equations involving six unknown average quantities, 
\ba
&&  0 =  - \frac{2}{\tau_r}  \langle  v^2  \rangle_+
  +   \frac{2 dD}{\tau_r^2}    -    \frac{1}{\tau} \langle v^2 \rangle_+     +     \frac{1}{\tau} \langle v^2 \rangle_-   -  \frac{2 \mu K }{\tau_r}  \langle   {\bf r}\cdot {\bf v}     \rangle_+  \nonumber\\ 
&&  0 =   -
\frac{2}{\tau_r}  \langle  v^2  \rangle_-    +    \frac{2 d D}{\tau_r^2}   -        \frac{1}{\tau} \langle v^2 \rangle_-     +     \frac{1}{\tau} \langle v^2 \rangle_+         \nonumber\\ 
&& 0 = \langle  v^2 \rangle_+     +   \frac{1}{\tau_r}  \langle   {\bf v} \cdot {\bf r}    \rangle_+  
- \frac{1}{\tau} \langle {\bf v}\cdot{\bf r} \rangle_+     +     \frac{1}{\tau} \langle {\bf v}\cdot{\bf r} \rangle_-          -   \frac{ \mu K }{\tau_r}  \langle r^2   \rangle_+                             \nonumber\\ 
&& 0 = \langle v^2 \rangle_-   -    \frac{1}{\tau_r}  \langle   {\bf v}  \cdot  {}\bf r \rangle_-   -  \frac{1}{\tau} \langle {\bf v}\cdot{\bf r} \rangle_-     +     \frac{1}{\tau} \langle {\bf v}\cdot{\bf r} \rangle_+  \nonumber\\ 
&& 0 = 2\langle {\bf v}\cdot {\bf r} \rangle_+   - \frac{1}{\tau} \langle r^2 \rangle_+     +     \frac{1}{\tau} \langle r^2 \rangle_-       \nonumber\\ 
&& 0 = 2\langle {\bf v}\cdot {\bf r} \rangle_-     -   \frac{1}{\tau} \langle r^2 \rangle_-     +     \frac{1}{\tau} \langle r^2 \rangle_+.  
\ea
After solving the above system of coupled equations, we get 
\be
\langle v^2 \rangle - \frac{d D}{\tau_r} =  -\frac{d \tau D \mu K  (\tau_r+\tau)}{\mu K \tau_r^2 \tau - 4 \tau_r^2 - 6 \tau_r \tau-2 \tau^2}, 
\label{eq:v2-taur}
\ee
which for the overdamped regime $\tau_r=0$ reduces to
\be
\langle v^2 \rangle - \frac{d D}{\tau_r}    =    \frac{d D \mu K }{2}.  
\ee
Since $\langle v^2\rangle_{eq} = \frac{d D}{\tau_r}$, and since $\langle v^2 \rangle$ is proportional to the 
kinetic energy, we can think of the above quantity as representing the excess kinetic energy that 
arises due to fluctuations of a harmonic trap.  

In this work we focus on the overdapmed regime, however, since in Eq. (\ref{eq:v2-taur}) we derived 
a formula for a system with inertia, $\tau_t>0$, we 

the quantity on the left-hand-side is NOT independent of $\tau$
(or $\alpha=1/\tau \mu K$).  It only becomes independent if we set $\tau_r=0$, that is, in the overdamped
regime.  Consequently, the result in Eq. (\ref{eq:EP}), which tells us that $\langle \dot q\rangle$
is independent of $\alpha$, is specific to overdamped dynamics.

%



\end{document}